\documentclass[11pt]{article}
\usepackage[hmargin={3.0cm,3.0cm},vmargin={3.0cm,3.0cm}]{geometry}

\usepackage{amsmath,amssymb,amsfonts}
\usepackage{braket}
\usepackage{mathtools}
\usepackage{amsthm}
\usepackage{todonotes}
\usepackage{booktabs}
\usepackage{multirow}
\usepackage{microtype}
\usepackage{subfig}
\usepackage{tikz}
\usetikzlibrary{arrows.meta}
\usetikzlibrary{decorations.markings}
\usetikzlibrary{shapes.geometric}

\usepackage{enumitem}

\usepackage{bm}
\usepackage[skip=2pt,font=small]{caption}

\usepackage{hyperref}
\hypersetup{
colorlinks,
linkcolor={blue!60!black},
citecolor={blue!90!black},
urlcolor={blue!90!black}}
\usepackage{cleveref}

\numberwithin{equation}{section}

\usepackage[titletoc,title]{appendix}

\newtheorem{theorem}{Theorem}[section]
\newtheorem{cor}[theorem]{Corollary}
\newtheorem{prop}[theorem]{Proposition}

\theoremstyle{definition}
\newtheorem{definition}[theorem]{Definition}
\theoremstyle{remark}
\newtheorem{remark}[theorem]{Remark}

\newcommand\hx{12pt} 
\newcommand\vs{1.7pt}

\tikzset{->-/.style={decoration={
  markings,
  mark=at position #1 with {\arrow{Latex}}},postaction={decorate}}}
\tikzset{->>-/.style={decoration={
  markings,
  mark=at position .5 with {\arrow{Latex[sep=10pt]Latex}}},postaction={decorate}}}
\tikzset{-<-/.style={decoration={
  markings,
  mark=at position #1 with {\arrow{Latex[reversed]}}},postaction={decorate}}}
\tikzset{-<<-/.style={decoration={
  markings,
  mark=at position .5 with {\arrow{Latex[reversed,sep=-10pt]Latex[reversed]}}},postaction={decorate}}}
  
\tikzset{-|-/.style={decoration={
  markings,
  mark=at position .51 with {\arrow{Bar}}},postaction={decorate}}}
\tikzset{-||-/.style={decoration={
  markings,
  mark=at position .49 with {\arrow{Bar[sep=-5pt] Bar}}},postaction={decorate}}}
\tikzset{-!-/.style={decoration={
  markings,
  mark=at position #1 with {\arrow{Rays[n=6,length=6pt]}}},postaction={decorate}}}
\tikzset{-!!-/.style={decoration={
  markings,
  mark=at position .51 with {\arrow{Bar[sep=1pt,length=4pt] Rays[n=6,length=8pt]}}},postaction={decorate}}}

\DeclarePairedDelimiter{\abs}{|}{|}
\newcommand{\R}{\mathbb{R}}
\newcommand{\N}{\mathbb{N}}
\newcommand{\Z}{\mathbb{Z}}
\newcommand{\Cn}{\mathbb{C}}

\newcommand{\x}{{\boldsymbol{x}}}
\newcommand{\y}{{\boldsymbol{y}}}
\newcommand{\al}{{{\alpha}}}
\newcommand{\bt}{{{\beta}}}
\newcommand{\gm}{{{\gamma}}}

\renewcommand{\imath}{{\mathrm{i}}}

\newcommand{\Pj}{\mathbb{CP}}

\newcommand{\A}[8]{A(#1;#2,#3,#4,#5;#6,#7,#8)}
\newcommand{\B}[8]{B(#1;#2,#3,#4,#5;#6,#7,#8)}
\newcommand{\C}[8]{C(#1;#2,#3,#4,#5;#6,#7,#8)}
\newcommand{\Cbar}[8]{\overline{C}(#1;#2,#3,#4,#5;#6,#7,#8)}

\newcommand{\xa}{x_a}
\newcommand{\xb}{x_b}
\newcommand{\xc}{x_e}
\newcommand{\xd}{x_d}
\newcommand{\xe}{x_f}
\newcommand{\xf}{x_c}
\newcommand{\Aa}{A.a}
\newcommand{\Ab}{A.b}
\newcommand{\Af}{A.c}
\newcommand{\Ad}{A.d}
\newcommand{\Ac}{A.e}
\newcommand{\Ae}{A.f}
\newcommand{\Ba}{B.a}
\newcommand{\Bb}{B.b}
\newcommand{\Bf}{B.c}
\newcommand{\Bd}{B.d}
\newcommand{\Bc}{B.e}
\newcommand{\Be}{B.f}
\newcommand{\Ca}{C.a}
\newcommand{\Cb}{C.b}
\newcommand{\Cf}{C.c}
\newcommand{\Cd}{C.d}
\newcommand{\Cc}{C.e}
\newcommand{\Ce}{C.f}

\newcommand{\hexCneq}[9]{\bm{C}^{(hex)}(#1,#2,#3,#4,#5,#6;#7,#8,#9)}
\newcommand{\hexA}[9]{\bm{A}^{(hex)}(#1,#2,#3,#4,#5,#6;#7,#8,#9)=0}
\newcommand{\hexC}[9]{\bm{C}^{(hex)}(#1,#2,#3,#4,#5,#6;#7,#8,#9)=0}

\newcommand{\sh}[1]{s(#1)}
\newcommand{\zz}[1]{z(#1)}
\newcommand{\shn}[2]{s_{#1 #2}}

\def\EXP{\textrm{{\large e}}}

\usepackage[backend=biber,sorting=nyt,maxnames=99,giveninits=true,doi=false,isbn=false,url=false,citestyle=numeric-comp]{biblatex}
\title{Algebraic entropy for hex systems}

\date{\today}

\usepackage{authblk}

\begin{document}
\author[1]{Giorgio Gubbiotti\footnote{Author to whom any correspondence should
be addressed.}}
\author[2]{Andrew P.~Kels}
\author[3]{Claude-M. Viallet}

\affil[1]{Universit\`a degli Studi di Milano, Dipartimento di Matematica
``Federigo Enriques'', Via Cesare Saldini 50, 20133, Milano, Italy, \& INFN
Sezione di Milano, Via Giovanni Celoria 16, 20133, Milano, Italy
    
\emph{email:} \texttt{giorgio.gubbiotti@unimi.it}}

\affil[2]{School of Mathematics and Statistics, The University of New South Wales, Sydney, NSW 2052, Australia

\emph{email:} \texttt{a.kels@unsw.edu.au}}

\affil[3]{LPTHE, UMR 7589 Centre National de la Recherche Scientifique \& 
  Sorbonne Universit\'e, 4 place Jussieu, 75252 Paris Cedex 05, France
  
 \emph{email:} \texttt{claude.viallet@upmc.fr}}

\maketitle

\begin{abstract}
  Hex systems were recently introduced \cite{Kels2022} as
  systems of equations defined on two-dimensional honeycomb lattices.
  We give a definition of algebraic entropy for such systems and use
  it to check the integrability of specific examples.
\end{abstract}

\section{Introduction}
The notion of integrability was introduced at the end of the 19th
century by Liouville in the context of classical mechanics, i.e.\ for
ordinary differential equations. Since then this notion has been
extended to many other classes of dynamical systems. What propelled
the ideas of integrability during the second half of the 20th century
was the discovery of elastic interactions of \emph{solitons} in the
Korteweg--De Vries equation, a partial differential equation appearing
in the theory of shallow waters, by Zabusky and
Kruskal~\cite{Zabusky1965}. Integrable systems are of special interest
because their solutions have non-trivial and fascinating
properties. Moreover, they are ``universal'', in the sense that almost
all integrable systems arise as the limiting form of many nonlinear
systems under appropriate scaling~\cite{Calogero1991}.

A further step was taken with the extension of the notion of integrability to
discrete systems (see for example~\cite{HietarintaJoshiNijhoff2016}).
Differential evolution equations then become difference equations, and partial
differential equations become for example ``quad equations'' defined on a
two-dimensional square lattice~\cite{BS2002}. For such systems the integrable
cases have been proficiently isolated through \emph{the consistency
approach}~\cite{DoliwaSantini1997,Nijhoff2001,ABS2003}. Consistency methods had
great success because of their algorithmic nature, and because in most of the
cases they yield a Lax pair~\cite{Nijhoff2002,BS2002}.  This
    is important because from a \emph{bona fide} Lax pair it is possible to
    construct infinitely many conservation
    laws~\cite{LeviWinternitzYamilov2022Book}, and build \emph{discrete
    $N$-soliton
solutions}~\cite{HietarintaZhang2009,Nijhoff_etal2009,Atkinson_etal2007,Atkinson_etal2008},
thus \emph{proving} integrability.  These last two properties must be checked
on a case-by-case basis because it has been discovered that sometimes the Lax
pair generated by the consistency method is not a bona fide one ~\cite{GSL_general,GSL_Gallipoli15},
in which case it cannot produce infinitely many conservation laws. This usually happens when the
associated quad equations are linearisable and/or Darboux
integrable~\cite{GSY_DarbouxI,GSY_DarbouxII,AdlerStartsev1999}.

Given a system of equations, one natural question arises: can we test if it
falls into the integrable category or not? One simple approach is the
calculation of the \emph{algebraic
entropy}~\cite{Veselov1992,FalquiViallet1993,HietarintaViallet1998,BellonViallet1999},
which measures the complexity of the system.  Two conditions are required: one
should be able  to define an evolution, and this evolution should be given by
birational transformations.  One evaluates the growth of the degrees of the
iterates under the evolution, in terms of generic initial conditions
(see~\Cref{sec:aedef}).  Exponential growth means non vanishing entropy, and
integrability is signalled by vanishing of the entropy (polynomial growth of
the sequence of degrees).  The special case of linear growth indicates that the
equations are linearisable.

The interest of this quick test is to easily discriminate integrable from
non-integrable cases, before embarking into a detailed study. 

In this paper we describe how to calculate the algebraic entropy and evaluate
it for a class of discrete systems, whose degrees of freedom take values on a
honeycomb lattice, called the \emph{hex systems}~\cite{Kels2022}.  A hex system
provides an important example of a system of equations which is defined in a
lattice different from the square lattice, and for which one can define a
birational evolution. The essential idea for this is that for hex systems one
considers a system of equations defined on each hexagonal cell, instead of just
a single equation.

The structure of this paper is as follows. In \Cref{sec:hex} we introduce
hex systems, discuss their discrete symmetries, and present a
list of hex systems obtained from a class of equations known as face-centered quad equations \cite{Kels2019}.
In \Cref{sec:aedef} we explain how to compute the algebraic entropy
for hex systems.
In \Cref{sec:examples} the  entropy is evaluated explicitly for
specific examples of hex systems defined in terms of face-centered quad
equations. The interested reader will find the
explicit expression of all the face-centered quad equations we
used in~\ref{app:equations}.
Finally, in \Cref{sec:concl} we give a quick outlook of our results in
comparison with previous ones, and present a list of open problems.

\section{Hex systems}
\label{sec:hex}
\begin{figure}[htb!]
    \centering
    \pgfmathsetmacro\tr{sqrt(3)/2}
    \begin{tikzpicture}[scale=2.0]

    \coordinate (A1) at (-1,0);
    \coordinate (A2) at (-1/2,{\tr});
    \coordinate (A3) at (1/2,{\tr});
    \coordinate (A4) at (1,0);
    \coordinate (A5) at (1/2,{-\tr});
    \coordinate (A6) at (-1/2,{-\tr});

    \draw (A1)--(A2) node[midway,above left]{$\gamma$};
    \draw (A2)--(A3) node[midway,above]{$\alpha$};
    \draw (A3)--(A4) node[midway,above right]{$\beta$};
    \draw (A4)--(A5) node[midway,below right]{$\gamma$};
    \draw (A5)--(A6) node[midway,below]{$\alpha$};
    \draw (A6)--(A1) node[midway,below left]{$\beta$};

    \fill (A1) circle (1.4pt) node[left]{$\xa$};
    \fill (A2) circle (1.4pt) node[above left]{$\xb$};
    \fill (A3) circle (1.4pt) node[above right]{$\xf$};
    \fill (A4) circle (1.4pt) node[right]{$\xd$};
    \fill (A5) circle (1.4pt) node[below right]{$\xc$};
    \fill (A6) circle (1.4pt) node[below left]{$\xe$};
    \end{tikzpicture}
    \caption{Elementary hexagonal cell}
    \label{fig:hexagonunitcell}
\end{figure}
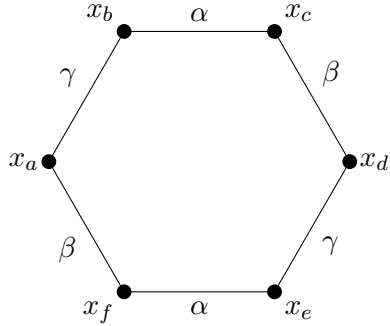

Consider the hexagon of Figure \ref{fig:hexagonunitcell}, with six
variables $x_a,x_b,x_c,x_d,x_e,x_f$ associated to its vertices, and
three parameters $\al,\bt,\gm$ associated to its edges.
Following~\cite{Kels2022}, one can in general construct a hex system
as the system of six equations
\begin{equation}
\begin{split}
A(x_a,x_b,x_c,x_d,x_e;\al,\bt,\gm)=0, \\
B(x_b,x_c,x_d,x_e,x_f;\al,\bt,\gm)=0, \\
C(x_c,x_d,x_e,x_f,x_a;\al,\bt,\gm)=0, \\
D(x_d,x_e,x_f,x_a,x_b;\al,\bt,\gm)=0, \\
E(x_e,x_f,x_a,x_b,x_c;\al,\bt,\gm)=0, \\
F(x_f,x_a,x_b,x_c,x_d;\al,\bt,\gm)=0,
\end{split}
\end{equation}
where $A,B,C,D,E,F$ are different functions, each depending linearly on at least one of its arguments, such that it is possible to solve at least one of the equations uniquely for each of the six variables on the hexagon.   
This requirement allows to define an evolution in the honeycomb lattice, which in
general depends on which of the six equations are used for the evolution.

To have an evolution independent of which equation is used, a consistency
condition may be imposed:
\begin{definition}
    A hex system is said to be {\emph{consistent}}, if after using two
    equations to solve for two variables at neighbouring vertices of the
    hexagon in terms of the other four, the remaining four equations are also
    satisfied on these solutions.
\end{definition}

This property was previously referred to as consistency-around-a-hexagon
\cite{Kels2022}.  For consistent hex systems we may define a birational
evolution that is independent of the equations chosen for the evolution (see
\Cref{sec:aedef}). Analogous notions of consistency for equations on
octahedrons have previously been formulated and studied in the context of
pluri-Lagrangian systems~\cite{BPS2014,BPS2016}.

We will limit ourselves to consistent hex systems constructed from
5-point face-centered quad equations which satisfy a recently
introduced property called consistency-around-a-face-centered-cube
\cite{Kels2019}.  Let $\A{x}{x_a}{x_b}{x_c}{x_d}{\al}{\bt}{\gm}=0$
denote a 5-point face-centered quad equation, where $A$ is a
polynomial that is multilinear in the four variables
$x_a,x_b,x_c,x_d$.  Such equations may be written in the form
\begin{equation}
    \A{x}{x_a}{x_b}{x_c}{x_d}{\al}{\bt}{\gm}
    =\sum_{k=0}^{\mu} x^{k} Q_k(x_a,x_b,x_c,x_d;\al,\bt,\gm)=0,
    \label{eq:Axmu}
\end{equation}
where each $Q_k$, $k=0,\ldots,\mu$, is a multilinear polynomial in the four
variables $x_a,x_b,x_c,x_d$, {\it i.e.}, each $Q_k=0$ is a regular quad
equation.

We will use two different types of face-centered quad equations, which
can be distinguished by their symmetries. 
Following~\cite{Kels2022}, we will call ``type A'' equations  $\A{x}{x_a}{x_b}{x_c}{x_d}{\al}{\bt}{\gm}=0$, the equations satisfying the following three symmetry conditions:
\begin{equation}\label{fcqsyms}
\begin{split}
\A{x}{x_a}{x_b}{x_c}{x_d}{\al}{\bt}{\gm}&=\sigma_1\A{x}{x_b}{x_a}{x_d}{x_c}{\al-\gm}{\bt-\gm}{-\gm}, \\
\A{x}{x_a}{x_b}{x_c}{x_d}{\al}{\bt}{\gm}&=\sigma_2\A{x}{x_c}{x_d}{x_a}{x_b}{\bt}{\al}{\gm}, \\
\A{x}{x_a}{x_b}{x_c}{x_d}{\al}{\bt}{\gm}&=\sigma_3\A{x}{x_d}{x_b}{x_c}{x_a}{-\al}{\al-\gm}{\bt-\al},
\end{split}
\end{equation}
where $\sigma_1,\sigma_2,\sigma_3\in\{-1,1\}$.

We will call ``type C'' equations
$\C{x}{x_a}{x_b}{x_c}{x_d}{\al}{\bt}{\gm}=0$, the equations satisfying the following symmetry
condition:
\begin{equation}
\begin{split}\label{typecsym}
\C{x}{x_a}{x_b}{x_c}{x_d}{\al}{\bt}{\gm}&=\sigma\C{x}{x_b}{x_a}{x_d}{x_c}{\al-\gm}{\bt-\gm}{-\gm}, 
\end{split}
\end{equation}
where $\sigma\in\{-1,1\}$ (this is equivalent to the first of \eqref{fcqsyms}).

Hex systems involve the variables and parameters associated to the
hexagon of Figure~\ref{fig:hexagonunitcell}.  The hex system
constructed from type A face-centered quad equations is defined as
\begin{equation}\label{Ahexdef}
\begin{alignedat}{2}
(\Aa)&&\qquad\A{\xa}{\xb}{\xf}{\xe}{\xc}{\bt}{\gm}{\al}=0, \\
(\Ab)&&\qquad\A{\xb}{\xf}{\xd}{\xa}{\xe}{\gm}{\al}{\bt}=0, \\
(\Af)&&\qquad\A{\xf}{\xd}{\xc}{\xb}{\xa}{\al}{\bt}{\gm}=0, \\
(\Ad)&&\qquad\A{\xd}{\xc}{\xe}{\xf}{\xb}{\bt}{\gm}{\al}=0, \\
(\Ac)&&\qquad\A{\xc}{\xe}{\xa}{\xd}{\xf}{\gm}{\al}{\bt}=0, \\
(\Ae)&&\qquad\A{\xe}{\xa}{\xb}{\xc}{\xd}{\al}{\bt}{\gm}=0.
\end{alignedat}
\end{equation}

The system of equations \eqref{Ahexdef} will be denoted by
$\hexA{\xa}{\xb}{\xf}{\xd}{\xc}{\xe}{\al}{\bt}{\gm}$, and referred to
as a \emph{hex system of type A}.  The hex system \eqref{Ahexdef} is
manifestly symmetric under rotations of the hexagon by $\pi/3$, given
by
\begin{equation}\label{rotsym}
(x_a,x_b,x_c,x_d,x_e,x_f)\to(x_b,x_c,x_d,x_e,x_f,x_a),\qquad (\al,\bt,\gm)\to(\bt,\gm,\al).
\end{equation}

Let $\C{x}{x_a}{x_b}{x_c}{x_d}{\al}{\bt}{\gm}=0$ and
$\Cbar{x}{x_a}{x_b}{x_c}{x_d}{\al}{\bt}{\gm}=0$ denote two type C
face-centered quad equations.  The hex system is then defined as
\begin{equation}\label{Chexdef}
\begin{alignedat}{2}
(\Ca)&&\qquad\Cbar{\xa}{\xe}{\xc}{\xb}{\xf}{\gm}{\bt}{\al}=0, \\
(\Cb)&&\qquad\C{\xb}{\xf}{\xd}{\xa}{\xe}{\gm}{\al}{\bt}=0, \\
(\Cf)&&\qquad\C{\xf}{\xb}{\xd}{\xa}{\xc}{\gm-\al}{-\al}{\bt-\al}=0, \\
(\Cd)&&\qquad\C{\xd}{\xf}{\xb}{\xc}{\xe}{\gm}{\bt}{\al}=0, \\
(\Cc)&&\qquad\Cbar{\xc}{\xe}{\xa}{\xd}{\xf}{\gm}{\al}{\bt}=0, \\
(\Ce)&&\qquad\Cbar{\xe}{\xc}{\xa}{\xd}{\xb}{\gm-\al}{-\al}{\bt-\al}=0.
\end{alignedat}
\end{equation}
The system of equations \eqref{Chexdef} will be denoted by
$\hexC{\xa}{\xb}{\xf}{\xd}{\xc}{\xe}{\al}{\bt}{\gm}$, and referred to
as a \emph{hex system of type C}.  

Unlike the type A system, the hex system \eqref{Chexdef} is not
invariant under the rotation of the hexagon by $\pi/3$ given by
\eqref{rotsym}. 
For type A face-centered quad equations which satisfy
the symmetries \eqref{fcqsyms}, the hex system of type C
\eqref{Chexdef} is equivalent to the hex system of type A
\eqref{Ahexdef}.  In this sense, the hex system of type C may be
regarded as a generalisation of the rotationally symmetric hex system
of type A.

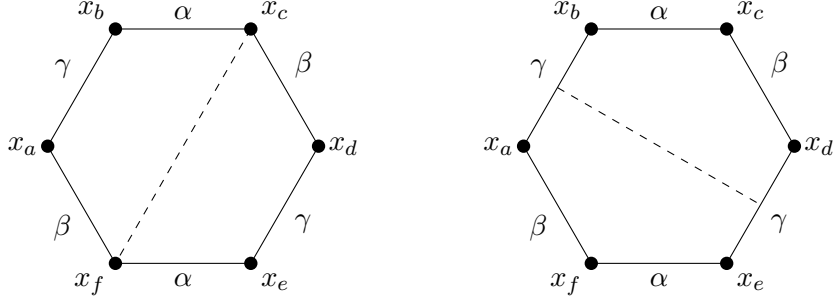
\begin{figure}[htb!]
\centering
\begin{tikzpicture}[scale=1.8]
\begin{scope}[xshift=100pt]
\pgfmathsetmacro\tr{sqrt(3)/2}
\coordinate (A1) at (-1,0);
\coordinate (A2) at (-1/2,{\tr});
\coordinate (A3) at (1/2,{\tr});
\coordinate (A4) at (1,0);
\coordinate (A5) at (1/2,{-\tr});
\coordinate (A6) at (-1/2,{-\tr});
\draw (A1)--(A2) node[midway,above left]{$\gamma$} coordinate[midway](A12);
\draw (A2)--(A3) node[midway,above]{$\alpha$};
\draw (A3)--(A4) node[midway,above right]{$\beta$};
\draw (A4)--(A5) node[midway,below right]{$\gamma$} coordinate[midway](A45);
\draw (A5)--(A6) node[midway,below]{$\alpha$};
\draw (A6)--(A1) node[midway,below left]{$\beta$};

\draw[dashed] (A12)--(A45);

\fill (A1) circle (1.4pt) node[left]{$\xa$};
\fill (A2) circle (1.4pt) node[above left]{$\xb$};
\fill (A3) circle (1.4pt) node[above right]{$\xf$};
\fill (A4) circle (1.4pt) node[right]{$\xd$};
\fill (A5) circle (1.4pt) node[below right]{$\xc$};
\fill (A6) circle (1.4pt) node[below left]{$\xe$};
\end{scope}

\begin{scope}[xshift=0pt]
\pgfmathsetmacro\tr{sqrt(3)/2}
\coordinate (A1) at (-1,0);
\coordinate (A2) at (-1/2,{\tr});
\coordinate (A3) at (1/2,{\tr});
\coordinate (A4) at (1,0);
\coordinate (A5) at (1/2,{-\tr});
\coordinate (A6) at (-1/2,{-\tr});

\draw (A1)--(A2) node[midway,above left]{$\gamma$} coordinate[midway](A12);
\draw (A2)--(A3) node[midway,above]{$\alpha$};
\draw (A3)--(A4) node[midway,above right]{$\beta$};
\draw (A4)--(A5) node[midway,below right]{$\gamma$} coordinate[midway](A45);
\draw (A5)--(A6) node[midway,below]{$\alpha$};
\draw (A6)--(A1) node[midway,below left]{$\beta$};

\draw[dashed] (A3)--(A6);

\fill (A1) circle (1.4pt) node[left]{$\xa$};
\fill (A2) circle (1.4pt) node[above left]{$\xb$};
\fill (A3) circle (1.4pt) node[above right]{$\xf$};
\fill (A4) circle (1.4pt) node[right]{$\xd$};
\fill (A5) circle (1.4pt) node[below right]{$\xc$};
\fill (A6) circle (1.4pt) node[below left]{$\xe$};
\end{scope}
\end{tikzpicture}
\caption{Reflections of a hexagon.}
\label{fig:d6syms}
\end{figure}

If $C$ and $\overline{C}$ are two distinct equations, then the type C hex system 
only satisfies the reflection symmetry:
\begin{equation}
\hexCneq{\xa}{\xb}{\xf}{\xd}{\xc}{\xe}{\al}{\bt}{\gm}
=\hexCneq{\xc}{\xd}{\xf}{\xb}{\xa}{\xe}{\bt}{\al}{\gm},
\end{equation}
which is indicated in the first diagram of \Cref{fig:d6syms}.

If $C=\overline{C}$, then the type C hex system 
also satisfies the reflection symmetry: 
\begin{equation}
\hexCneq{\xa}{\xb}{\xf}{\xd}{\xc}{\xe}{\al}{\bt}{\gm}
=\hexCneq{\xb}{\xa}{\xe}{\xc}{\xd}{\xf}{\bt}{\al}{\gm}, \\
\end{equation}
which is indicated in the second diagram of \Cref{fig:d6syms}.

Finally, the following rotation of the hexagon by $\pi$
\begin{equation}
\xa\leftrightarrow\xd,\quad \xb\leftrightarrow\xc, \quad \xf\leftrightarrow\xe,
\end{equation}
has the effect of exchanging the equations $(\Ca)\leftrightarrow(\Cd)$,
$(\Cb)\leftrightarrow(\Cc)$, $(\Cf)\leftrightarrow(\Ce)$, and simultaneously
exchanging $C\leftrightarrow\overline{C}$, and thus the type C hex system is
invariant under rotation of the hexagon by $\pi$, if $C=\overline{C}$. 

Examples of consistent hex systems of type A and type C were previously given
in~\cite{Kels2022}. Examples of consistent hex systems of type A can be
constructed from each of the type A face-centered quad equations listed
in~\ref{app:equations}. Examples of consistent hex systems of type C can be
constructed from each of the pairs of type C face-centered quad equations
listed in Table~\ref{tab:typeCpolytope}, with the explicit equations listed
in~\ref{app:equations}. Such systems of consistent equations were previously
studied in the context of octahedron relations obtained from pluri-Lagrangian
systems~\cite{BS2009,BPS2014,BPS2016}, where in~\cite{Kels2022} this octahedron
structure is implicitly inherited from their derivation from the star-star
relation, an integrability condition for lattice models of statistical
mechanics related to the Yang-Baxter
equation~\cite{BaxterBook,Baxter1997,BazhanovSergeev2012}.
\ref{app:equations} also lists examples of so-called type B equations, which
are equations that satisfy the first two symmetries of~\eqref{fcqsyms}, but it
is not known if it is possible to use these equations to construct consistent
hex systems.

Contrarily to what happens for quad equations for which the local condition on
a cell is given by \emph{one equation}, for hex systems we will need \emph{two
equations} per cell to define the model.

\begin{table}[htb!]
\centering

\begin{tabular}{cc}
        \toprule
 $C$ & $\overline{C}$ 
 \\
 \midrule
$C3_{(1/2;\,1/2;\,0)}$  &  
$C3_{(1/2;\,0;\,1/2)}$  
\\[0.11cm]
$C3_{(1/2;\,0;\,1/2)}$  &  
$C3_{(1/2;\,1/2;\,0)}$  
\\[0.11cm]
$C3_{(1;\,0;\,0)}$  &  
\textrm{same as} $C$ 
\\[0.11cm]
$C3_{(0;\,0;\,0)}$  &  
\textrm{same as} $C$ 
\\[0.11cm]
$C2_{(1;\,1;\,0)}$  &  
$C2_{(1;\,0;\,1)}$ 
\\[0.11cm]
$C2_{(1;\,0;\,1)}$  &  
$C2_{(1;\,1;\,0)}$  
\\[0.11cm]
$C2_{(1;\,0;\,0)}$ &
\textrm{same as} $C$ 
\\[0.11cm]
$C1_{(1)}$  &  
$C2_{(0;\,0;\,0)}$ 
\\[0.11cm]
$C2_{(0;\,0;\,0)}$ &
$C1_{(1)}$  
\\[0.11cm]
$C1_{(0)}$ &
\textrm{same as} $C$ 
\\[0.05cm]
\bottomrule
\end{tabular}
\caption{Examples of pairs of type C equations that form consistent hex systems.}
\label{tab:typeCpolytope}
\end{table}

\section{Algebraic entropy for hex systems}
\label{sec:aedef}

\subsection{Initial conditions and evolution}

In order to associate an entropy to a model, we need to define some evolution
map and evaluate the sequence of degrees of its successive iterates, once
initial conditions are given, see for instance
\cite{Tremblay2001,Viallet2006,GubKelsEntropy,Gubbiotti2023DecioMemorial,Hietarinta2023Bous}.
At the cell level, there are six possible directions of evolution,
corresponding to the choices of a pair of variables with respect to which we
solve the system \eqref{Ahexdef} or \eqref{Chexdef}. So, in analogy with the
quad and face-centered quad cases we name the directions of evolution (denoted
$\Delta$), north-east (NE), north-west (NW), north (N), \dots, with obvious
denomination. 

\begin{figure}[htb!]
    \centering
    \pgfmathsetmacro\tr{sqrt(3)/2}
    
   \begin{tikzpicture}[scale=2.0]

    \coordinate (A1) at (-1,0);
    \coordinate (A2) at (-1/2,{\tr});
    \coordinate (A3) at (1/2,{\tr});
    \coordinate (A4) at (1,0);
    \coordinate (A5) at (1/2,{-\tr});
    \coordinate (A6) at (-1/2,{-\tr});

    \draw (A1)--(A2); 
    \draw (A2)--(A3); 
    \draw (A3)--(A4); 
    \draw (A4)--(A5); 
    \draw (A5)--(A6); 
    \draw (A6)--(A1); 


    \fill[red] (A1) circle (1.4pt) node[red, left]{$\xa$};
    \fill (A2) circle (1.4pt) node[above left]{$\xb$};
    \fill[red] (A3) circle (1.4pt) node[red, above right]{$\xf$};
    \fill[red] (A4) circle (1.4pt) node[red,right]{$\xd$};
    \fill (A5) circle (1.4pt) node[below right]{$\xc$};
    \fill[red] (A6) circle (1.4pt) node[red,below left]{$\xe$};

\draw[red,ultra thick, ->] (-.75,-1/2*\tr) -- (.75,1/2*\tr); 

   \end{tikzpicture}
   \caption{North-east evolution on one cell}
    \label{fig:one-cell}
   \end{figure}
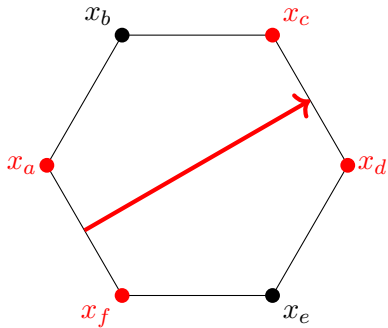

\Cref{fig:one-cell} pictures the NE evolution at the level of
one cell: the values of $x_c$ and $x_d$ are calculated in terms of
$x_a$ and $x_f$, with the other two variables $x_b$ and $x_e$ being
spectators. It is important to notice that we define in this way a
{\em{birational evolution}} $(x_a,x_f) \mapsto (x_c,x_d)$.  This will
be the case for {\em all the models we consider and all directions of
  evolution}: we have two relations between the values $x_v$ at the
six vertices $ v= a,b,c,d,e,f$, {\em and these relations can be solved
  rationally for all pairs of adjacent  vertices in terms of the four remaining
  ones}.

For each direction $\Delta$ we solve the system \eqref{Ahexdef} or
\eqref{Chexdef} with respect to a pair of distinguished vertices
$D_\Delta$, as indicated in Table \ref{tab:Dist}.

\begin{table}[h]
    \centering
    \begin{tabular}{cc}
        \toprule
        \textbf{Direction $\Delta$} & \textbf{Distinguished vertices $D_\Delta$} 
            \\
        \midrule
        NE & $ c$, $d$
        \\
        N & $b$, $c$ 
        \\
        NW & $a$, $b$
        \\
        SW & $a$, $f$
        \\
        S & $e$, $f$
        \\
        SE & $d$, $e$
        \\
        \bottomrule
      \end{tabular}
   \caption{Distinguished vertices  for the various     directions of evolution.}
    \label{tab:Dist}
\end{table}

We note that the situation we are considering is different from the one
of \cite{vanderKamp2009}, because for each cell we use two equations to evaluate the
dynamical variables at two \emph{adjacent sites} in terms of the other four.

We choose initial conditions on a staircase, an example of which is shown in \Cref{fig: ne-evol} for evolutions in the NE and SW directions.  It is important to notice
that this choice of initial conditions allows to cover the entire
plane with the evolution and its inverse.
Of course, in order to evaluate explicit degrees, we will start from a
{\em finite staircase}, allowing us to calculate
iterates on a triangular domain.  If we wish to perform $N$ iterations we need to specify $4N$ initial conditions.

\renewcommand\hx{12pt} 
\renewcommand\vs{1.7pt}
\pgfmathsetmacro\cs{cos(30)} 

\begin{figure}[htb!]
    \centering
             {
        \begin{tikzpicture}[scale=1.4]

\foreach \y in {0,1,2,3,4,5}{
\foreach \x in {0,1,2,3,4}{

\pgfmathsetmacro\tmpo{\y-0} 

\ifnum\x<\tmpo{

\ifnum\y=5{

\draw (3*\x*\hx/2 -3*\y*\hx,2*8*\hx*\cs-3*\x*\hx*\cs) --
(3*\x*\hx/2 -3*\y*\hx-\hx/2,2*8*\hx*\cs-3*\x*\hx*\cs-\hx*\cs); 
\draw (3*\x*\hx/2 -3*\y*\hx-\hx/2,2*8*\hx*\cs-3*\x*\hx*\cs-\hx*\cs) -- 
(3*\x*\hx/2 -3*\y*\hx,2*8*\hx*\cs-3*\x*\hx*\cs-2*\hx*\cs);
\draw (3*\x*\hx/2 -3*\y*\hx,2*8*\hx*\cs-3*\x*\hx*\cs-2*\hx*\cs) --
(3*\x*\hx/2 -3*\y*\hx+\hx,2*8*\hx*\cs-3*\x*\hx*\cs-2*\hx*\cs);

\filldraw[fill=black,draw=black] (3*\x*\hx/2 -3*\y*\hx,2*8*\hx*\cs-3*\x*\hx*\cs) circle (\vs);
\filldraw[fill=black,draw=black] (3*\x*\hx/2 -3*\y*\hx-\hx/2,2*8*\hx*\cs-3*\x*\hx*\cs-\hx*\cs) circle (\vs);
\filldraw[fill=black,draw=black] (3*\x*\hx/2 -3*\y*\hx,2*8*\hx*\cs-3*\x*\hx*\cs-2*\hx*\cs) circle (\vs);
\filldraw[fill=black,draw=black] (3*\x*\hx/2 -3*\y*\hx+\hx,2*8*\hx*\cs-3*\x*\hx*\cs-2*\hx*\cs) circle (\vs);

}\fi

\ifnum\y=4{

\draw (3*\x*\hx/2 -3*\y*\hx+\hx-3*5*\hx/2+2*\y*\hx/2+\hx/2,2*8*\hx*\cs-3*\x*\hx*\cs-6*\hx*\cs+\y*\hx*\cs) --
(3*\x*\hx/2 -3*\y*\hx+3*\hx/2-3*5*\hx/2+2*\y*\hx/2+\hx/2,2*8*\hx*\cs-3*\x*\hx*\cs-\hx*\cs-6*\hx*\cs+\y*\hx*\cs);

}\fi

\draw (3*\x*\hx/2 -3*\y*\hx-3*5*\hx/2+3*\y*\hx/2,2*8*\hx*\cs-3*\x*\hx*\cs-5*\hx*\cs+\y*\hx*\cs) --
(3*\x*\hx/2 -3*\y*\hx+\hx-3*5*\hx/2+3*\y*\hx/2,2*8*\hx*\cs-3*\x*\hx*\cs-5*\hx*\cs+\y*\hx*\cs);
\draw (3*\x*\hx/2 -3*\y*\hx+\hx-3*5*\hx/2+3*\y*\hx/2,2*8*\hx*\cs-3*\x*\hx*\cs-5*\hx*\cs+\y*\hx*\cs) --
(3*\x*\hx/2 -3*\y*\hx+3*\hx/2-3*5*\hx/2+3*\y*\hx/2,2*8*\hx*\cs-3*\x*\hx*\cs-\hx*\cs-5*\hx*\cs+\y*\hx*\cs);
\draw (3*\x*\hx/2 -3*\y*\hx+3*\hx/2-3*5*\hx/2+3*\y*\hx/2,2*8*\hx*\cs-3*\x*\hx*\cs-\hx*\cs-5*\hx*\cs+\y*\hx*\cs)  --
(3*\x*\hx/2 -3*\y*\hx+\hx-3*5*\hx/2+3*\y*\hx/2,2*8*\hx*\cs-3*\x*\hx*\cs-2*\hx*\cs-5*\hx*\cs+\y*\hx*\cs);

\filldraw[fill=white,draw=black] (3*\x*\hx/2 -3*\y*\hx+\hx-3*5*\hx/2+3*\y*\hx/2,2*8*\hx*\cs-3*\x*\hx*\cs-5*\hx*\cs+\y*\hx*\cs) circle (\vs);
\filldraw[fill=white,draw=black] (3*\x*\hx/2 -3*\y*\hx+3*\hx/2-3*5*\hx/2+3*\y*\hx/2,2*8*\hx*\cs-3*\x*\hx*\cs-\hx*\cs-5*\hx*\cs+\y*\hx*\cs) circle (\vs);

\ifnum\y<5{
\draw 
(3*\x*\hx/2 -15*\hx,2*8*\hx*\cs-3*\x*\hx*\cs-0*\hx*\cs-10*\hx*\cs+2*\y*\hx*\cs)--
(3*\x*\hx/2 -15*\hx-\hx/2,2*8*\hx*\cs-3*\x*\hx*\cs-\hx*\cs-10*\hx*\cs+2*\y*\hx*\cs)--
(3*\x*\hx/2 -15*\hx,2*8*\hx*\cs-3*\x*\hx*\cs-2*\hx*\cs-10*\hx*\cs+2*\y*\hx*\cs)--
(3*\x*\hx/2 -14*\hx,2*8*\hx*\cs-3*\x*\hx*\cs-2*\hx*\cs-10*\hx*\cs+2*\y*\hx*\cs);
\filldraw[fill=white,draw=black] (3*\x*\hx/2 -15*\hx-\hx/2,2*8*\hx*\cs-3*\x*\hx*\cs-\hx*\cs-10*\hx*\cs+2*\y*\hx*\cs) circle (\vs);
\filldraw[fill=white,draw=black] (3*\x*\hx/2 -15*\hx,2*8*\hx*\cs-3*\x*\hx*\cs-2*\hx*\cs-10*\hx*\cs+2*\y*\hx*\cs) circle (\vs);
}\fi

}\fi

}

}

        \end{tikzpicture}
    }
             \caption{A staircase of initial conditions (black sites) and calculable values (white sites).}
    \label{fig: ne-evol}
\end{figure}
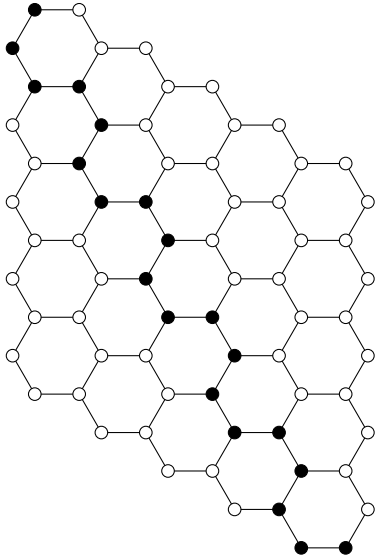

\begin{figure}[htb!]
    \centering
    \begin{tikzpicture}[x=6.8mm,y=3.85mm,scale=1.0]
    \tikzset{box/.style={regular polygon,
        regular polygon sides=6,
        minimum size=9mm,
        inner sep=0mm,
        outer sep=0mm,
        rotate=0,
        draw
        }
    }

    \foreach \i in {0,...,5} 
        \foreach \j in {-1,...,5} {
                \node[box] at (2*\i,-2*\j) {};
                \node[box] at (2*\i-1,-2*\j-1) {};
    }

    \node[box,fill=green!15] at (-1,1) {\tiny{$[1,1]$}};
    \node[box,fill=green!15] at (0,-2) {\tiny{$[1,2]$}};
    \node[box,fill=green!15] at (1,-5) {\tiny{$[1,3]$}};
    \node[box,fill=green!15] at (2,-8) {\tiny{$[1,4]$}};
    \node[box,fill=green!15] at (3,-11) {\tiny{$[1,5]$}};
    \node[box,fill=green!15] at (0,0) {\tiny{$[2,1]$}};
    \node[box,fill=green!15] at (1,-3) {\tiny{$[2,2]$}};
    \node[box,fill=green!15] at (2,-6) {\tiny{$[2,3]$}};
    \node[box,fill=green!15] at (3,-9) {\tiny{$[2,4]$}};
    \node[box,fill=green!15] at (1,-1) {\tiny{$[3,1]$}};
    \node[box,fill=green!15] at (2,-4) {\tiny{$[3,2]$}};
    \node[box,fill=green!15] at (3,-7) {\tiny{$[3,3]$}};
    \node[box,fill=green!15] at (2,-2) {\tiny{$[4,1]$}};
    \node[box,fill=green!15] at (3,-5) {\tiny{$[4,2]$}};
    \node[box,fill=green!15] at (3,-3) {\tiny{$[5,1]$}};

    \node[box,fill=magenta!15] at (10,2) {\tiny{$[1,1]$}};
    \node[box,fill=magenta!15] at (9,-1) {\tiny{$[1,2]$}};
    \node[box,fill=magenta!15] at (8,-4) {\tiny{$[1,3]$}};
    \node[box,fill=magenta!15] at (7,-7) {\tiny{$[1,4]$}};
    \node[box,fill=magenta!15] at (6,-10) {\tiny{$[1,5]$}};
    \node[box,fill=magenta!15] at (9,1) {\tiny{$[2,1]$}};
    \node[box,fill=magenta!15] at (8,-2) {\tiny{$[2,2]$}};
    \node[box,fill=magenta!15] at (7,-5) {\tiny{$[2,3]$}};
    \node[box,fill=magenta!15] at (6,-8) {\tiny{$[2,4]$}};
    \node[box,fill=magenta!15] at (8,0) {\tiny{$[3,1]$}};
    \node[box,fill=magenta!15] at (7,-3) {\tiny{$[3,2]$}};
    \node[box,fill=magenta!15] at (6,-6) {\tiny{$[3,3]$}};
    \node[box,fill=magenta!15] at (7,-1) {\tiny{$[4,1]$}};
    \node[box,fill=magenta!15] at (6,-4) {\tiny{$[4,2]$}};
    \node[box,fill=magenta!15] at (6,-2) {\tiny{$[5,1]$}};


\begin{scope}[yshift=-180,xshift=20]


    \foreach \i in {0,...,4} 
        \foreach \j in {-1,...,5} {
                \node[box] at (2*\i,-2*\j) {};
                \node[box] at (2*\i-1,-2*\j-1) {};
    }

    \node[box,fill=green!15] at (-1,-11) {\tiny{$[1,1]$}};
    \node[box,fill=green!15] at (1,-11) {\tiny{$[1,2]$}};
    \node[box,fill=green!15] at (3,-11) {\tiny{$[1,3]$}};
    \node[box,fill=green!15] at (5,-11) {\tiny{$[1,4]$}};
    \node[box,fill=green!15] at (7,-11) {\tiny{$[1,5]$}};
    \node[box,fill=green!15] at (0,-10) {\tiny{$[2,1]$}};
    \node[box,fill=green!15] at (2,-10) {\tiny{$[2,2]$}};
    \node[box,fill=green!15] at (4,-10) {\tiny{$[2,3]$}};
    \node[box,fill=green!15] at (6,-10) {\tiny{$[2,4]$}};
    \node[box,fill=green!15] at (1,-9) {\tiny{$[3,1]$}};
    \node[box,fill=green!15] at (3,-9) {\tiny{$[3,2]$}};
    \node[box,fill=green!15] at (5,-9) {\tiny{$[3,3]$}};
    \node[box,fill=green!15] at (2,-8) {\tiny{$[4,1]$}};
    \node[box,fill=green!15] at (4,-8) {\tiny{$[4,2]$}};
    \node[box,fill=green!15] at (3,-7) {\tiny{$[5,1]$}};

    \node[box,fill=magenta!15] at (8,2) {\tiny{$[1,1]$}};
    \node[box,fill=magenta!15] at (6,2) {\tiny{$[1,2]$}};
    \node[box,fill=magenta!15] at (4,2) {\tiny{$[1,3]$}};
    \node[box,fill=magenta!15] at (2,2) {\tiny{$[1,4]$}};
    \node[box,fill=magenta!15] at (0,2) {\tiny{$[1,5]$}};
    \node[box,fill=magenta!15] at (7,1) {\tiny{$[2,1]$}};
    \node[box,fill=magenta!15] at (5,1) {\tiny{$[2,2]$}};
    \node[box,fill=magenta!15] at (3,1) {\tiny{$[2,3]$}};
    \node[box,fill=magenta!15] at (1,1) {\tiny{$[2,4]$}};
    \node[box,fill=magenta!15] at (6,0) {\tiny{$[3,1]$}};
    \node[box,fill=magenta!15] at (4,0) {\tiny{$[3,2]$}};
    \node[box,fill=magenta!15] at (2,0) {\tiny{$[3,3]$}};
    \node[box,fill=magenta!15] at (5,-1) {\tiny{$[4,1]$}};
    \node[box,fill=magenta!15] at (3,-1) {\tiny{$[4,2]$}};
    \node[box,fill=magenta!15] at (4,-2) {\tiny{$[5,1]$}};


\end{scope}



\begin{scope}[yshift=-360]


    \foreach \i in {0,...,4}{
    \node[box] at (2*\i-1,1) {};
    }

    \foreach \i in {0,...,5} 
        \foreach \j in {-1,...,5} {
                \node[box] at (2*\i,-2*\j) {};
                \node[box] at (2*\i-1,-2*\j-1) {};
    }

    \node[box,fill=green!15] at (3,1) {\tiny{$[1,1]$}};
    \node[box,fill=green!15] at (2,-2) {\tiny{$[1,2]$}};
    \node[box,fill=green!15] at (1,-5) {\tiny{$[1,3]$}};
    \node[box,fill=green!15] at (0,-8) {\tiny{$[1,4]$}};
    \node[box,fill=green!15] at (-1,-11) {\tiny{$[1,5]$}};
    \node[box,fill=green!15] at (3,-1) {\tiny{$[2,1]$}};
    \node[box,fill=green!15] at (2,-4) {\tiny{$[2,2]$}};
    \node[box,fill=green!15] at (1,-7) {\tiny{$[2,3]$}};
    \node[box,fill=green!15] at (0,-10) {\tiny{$[2,4]$}};
    \node[box,fill=green!15] at (3,-3) {\tiny{$[3,1]$}};
    \node[box,fill=green!15] at (2,-6) {\tiny{$[3,2]$}};
    \node[box,fill=green!15] at (1,-9) {\tiny{$[3,3]$}};
    \node[box,fill=green!15] at (3,-5) {\tiny{$[4,1]$}};
    \node[box,fill=green!15] at (2,-8) {\tiny{$[4,2]$}};
    \node[box,fill=green!15] at (3,-7) {\tiny{$[5,1]$}};

    \node[box,fill=magenta!15] at (6,2) {\tiny{$[1,1]$}};
    \node[box,fill=magenta!15] at (7,-1) {\tiny{$[1,2]$}};
    \node[box,fill=magenta!15] at (8,-4) {\tiny{$[1,3]$}};
    \node[box,fill=magenta!15] at (9,-7) {\tiny{$[1,4]$}};
    \node[box,fill=magenta!15] at (10,-10) {\tiny{$[1,5]$}};
    \node[box,fill=magenta!15] at (6,0) {\tiny{$[2,1]$}};
    \node[box,fill=magenta!15] at (7,-3) {\tiny{$[2,2]$}};
    \node[box,fill=magenta!15] at (8,-6) {\tiny{$[2,3]$}};
    \node[box,fill=magenta!15] at (9,-9) {\tiny{$[2,4]$}};
    \node[box,fill=magenta!15] at (6,-2) {\tiny{$[3,1]$}};
    \node[box,fill=magenta!15] at (7,-5) {\tiny{$[3,2]$}};
    \node[box,fill=magenta!15] at (8,-8) {\tiny{$[3,3]$}};
    \node[box,fill=magenta!15] at (6,-4) {\tiny{$[4,1]$}};
    \node[box,fill=magenta!15] at (7,-7) {\tiny{$[4,2]$}};
    \node[box,fill=magenta!15] at (6,-6) {\tiny{$[5,1]$}};
    

\end{scope}
\end{tikzpicture}
\caption{Indexing for evolution in the NE (top, lean green), NW (top, magenta), N (middle, lean green), S (middle, magenta), SE (bottom, lean green), and SW (bottom, magenta) directions.}
\label{fig:indexing}
\end{figure}
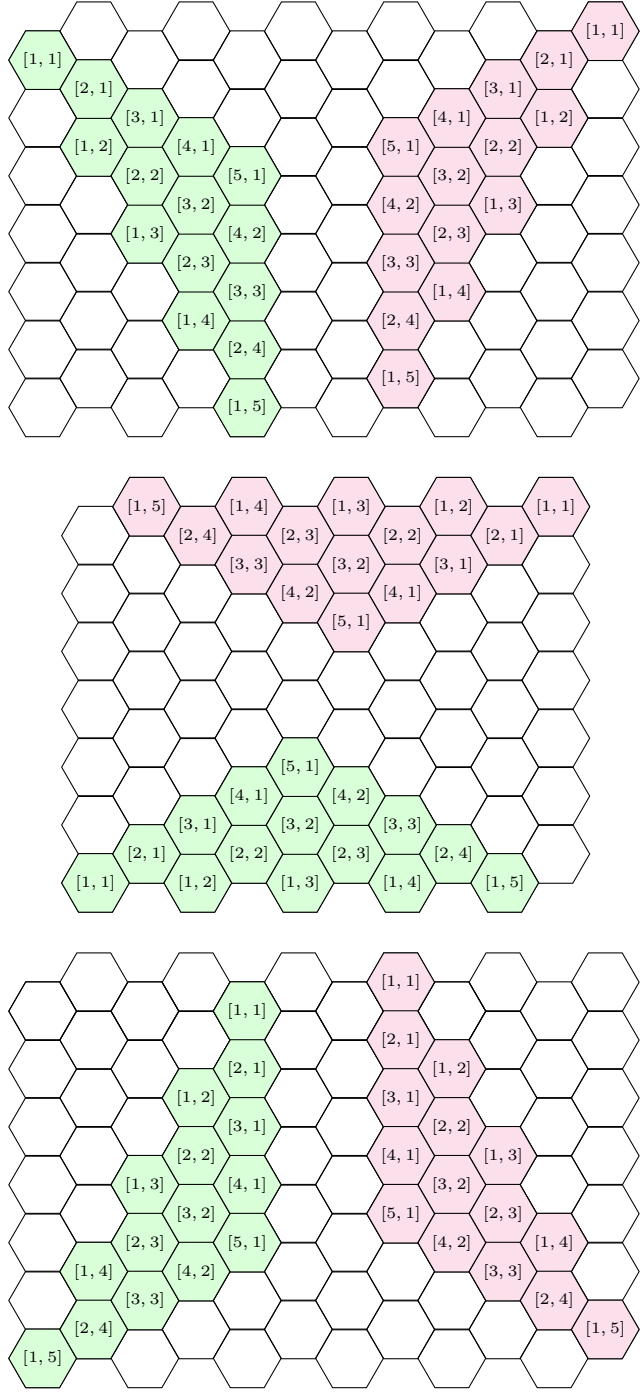

We attribute to each cell a double index $[n,j]$, where $1\leq n\leq N$ and $1\leq j\leq N-n+1$. The first index denotes the level of iteration, and the second index labels the position of the cell for a given level. Figure \ref{fig:3levit} shows the indexing of cells for evolution in the NE direction (indicated by the arrow), and Figure \ref{fig:indexing} 
    illustrates the indexing for each of the six different 
    directions of evolution for $N=5$.

    \begin{figure}[htb!]
        \centering
        \begin{tikzpicture}[scale=1.5]

        \pgfmathsetmacro\cs{cos(60)} 
        \pgfmathsetmacro\ss{sin(60)} 

        \draw (0,0) -- (1,0) -- (1+\cs,-\ss)--(2+\cs,-\ss) -- (2+2*\cs,-2*\ss)-- (2+\cs,-3*\ss)
        -- (2+2*\cs,-4*\ss) -- (2+\cs,-5*\ss) -- (1+\cs,-5*\ss) -- (1,-4*\ss) -- (0,-4*\ss) -- 
        (-\cs,-3*\ss) -- (0,-2*\ss) -- (-\cs,-\ss) -- cycle;
        \draw (1+\cs,-\ss) -- (1,-2*\ss) -- (0,- 2*\ss);
        \draw (1,-2*\ss) -- (1+\cs,- 3*\ss) -- (2+\cs,- 3*\ss);
        \draw (1+\cs,- 3*\ss) -- (1,- 4*\ss);
\draw[red,ultra thick, ->] (-.5,-3.5) -- (1+5*\cs,-.5-\ss); 
        \node at (\cs,-\ss) {$[n+1,j]$};
        \node at (\cs,-3*\ss) {$[n,j+1]$};
        \node at (1+2*\cs,-2*\ss) {$[n+2,j]$};
        \node at (1+2*\cs,-4*\ss) {$[n+1,j+1]$};
    \end{tikzpicture}
    \caption{Indexing of cells}
    \label{fig:3levit}
    \end{figure}
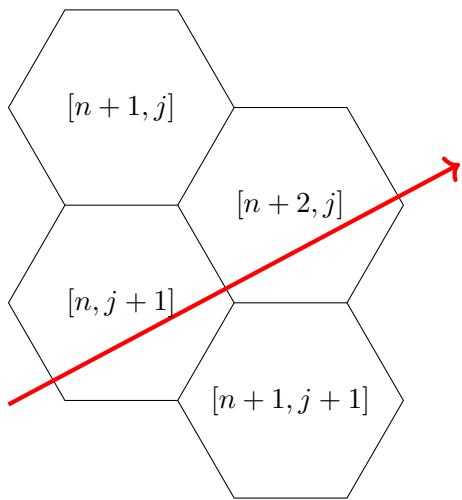

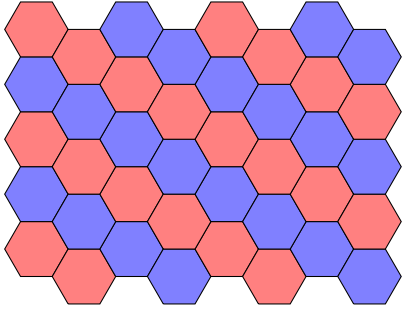
\begin{figure}[htb!]
    \centering
    \begin{tikzpicture}[rotate=0,scale=1]

        \newcommand\cola{blue!50!white}
        \newcommand\colb{red!50!white}

        \foreach \y in {0,1,2,3,4}{
        \foreach \x in {0,1,2,3}{
        \ifodd\inteval{\x+\y}{
        \filldraw[draw=black,fill=\cola] (3*\x*\hx,2*\y*\hx*\cs) -- ++(0:\hx) -- 
        ++(60:\hx) -- ++(120:\hx) -- ++(180:\hx) -- ++(240:\hx) -- ++(300:\hx)
        ;
        \filldraw[draw=black,fill=\cola] (3*\x*\hx+3*\hx/2,2*\y*\hx*\cs-\hx*\cs) -- 
        ++(0:\hx) -- ++(60:\hx) -- ++(120:\hx) -- ++(180:\hx) -- 
        ++(240:\hx) -- ++(300:\hx);
        }\fi
        \ifodd\inteval{\x+\y+1}{
        \filldraw[draw=black,fill=\colb] (3*\x*\hx,2*\y*\hx*\cs) -- ++(0:\hx) -- ++(60:\hx) -- ++(120:\hx) -- ++(180:\hx) -- ++(240:\hx) -- ++(300:\hx)
        ;
        \filldraw[draw=black,fill=\colb] (3*\x*\hx+3*\hx/2,2*\y*\hx*\cs-\hx*\cs) 
        -- ++(0:\hx) -- ++(60:\hx) -- ++(120:\hx) -- ++(180:\hx) -- 
        ++(240:\hx) -- ++(300:\hx);
        }\fi
        }
        }

    \end{tikzpicture}
    \caption{Bipartite lattice   rendered  as a blue-and-red lattice.}
    \label{fig:bipartite}
\end{figure}

In the case of hex systems of type A, the evolution at the level of the
cell does not depend on the location of the cell. In the
case of the hex systems of type C, sometimes different equations should be used throughout the lattice.  The evolution in those cases depends on the position on 
the lattice, alternating between two values, see \Cref{fig:bipartite}.
We call this case a \emph{bipartite system}.  This is analogous
to the \emph{black-and-white lattice} for quad equations
\cite{HietarintaViallet2012,ABS2009,Xenitidis2009}.

For our specific computations, we will use the $C$ equation for type C hex systems on the blue hexagons, and the $\bar{C}$ equation on the red hexagons. 
In that case, notice that for each direction $\Delta$, the forward evolution on the
blue hexagon coincides with the backward evolution on the red hexagons, which follows from the symmetries of hex systems of type C presented in Section \ref{sec:hex}. 



\subsection{Definitions}

For each cell with index $[n,j]$ ($ 1\leq n \leq N$, $ 1\leq j \leq N-n+1$), an
explicit calculation provides the values of $x_{v},\; v\in D_{v}$, as rational
functions of the initial data. The number of initial data is $4N$ since in each
cell we have $4$ initial data. The initial belong to the
complex projective space $\mathcal{X}=\Pj^{4N}$. In homogeneous projective
coordinates the numerator and the denominator of the $x_{v}$ have the same
degree, which we denote by $d_{v}^\Delta[n,j],\; v\in D_{v}$, and $\Delta = NE,
NW, SE, SW, N, S $.

\begin{definition}
Following \cite{BellonViallet1999} we define the algebraic entropy for
the vertex ${v}$ in the direction $\Delta$ (with $v \in D_\Delta$) as
\begin{equation}
  {\cal{S}}_{v}^\Delta = \lim_{n\to\infty}{\frac{1}{n} \log (d_{v}^\Delta [n,1])}
\end{equation}
\end{definition}

\bigskip
\noindent
In the example of the north-east evolution we thus define two meaningful
algebraic entropies ${\cal{S}}_c^{NE}$ and ${\cal{S}}_d^{NE}$ measuring
the growth of the degrees $d_c^{NE}$ and $d_d^{NE}$, as the evolution
goes on.

\noindent
We may then define a {\em directional  entropy} by
\begin{equation}
   {\cal{S}}^\Delta = \max_{ v \in D_\Delta} ({\cal{S}}_v^\Delta) 
\end{equation}
as well as a  {\em global entropy}
\begin{equation}
{\cal{S}} = \max_{\Delta} ( {\cal{S}}^\Delta) 
\end{equation}

\begin{definition}
    A hex system is \emph{globally integrable} if its global algebraic
    entropy $\cal{S}$ vanishes.
\end{definition}

We may also consider a finer characterisation of integrability:

\begin{definition}
  A hex system is {\em directionally integrable in the direction
    $\Delta$} if its directional entropy ${\cal{S}}^\Delta$
  vanishes.
\end{definition}

\subsection{Calculation}

We should in
principle know the entire sequence of degrees of the iterates to evaluate the algebraic entropy. 
The remarkable fact is that we can guess it from a finite part of the
sequence. The fundamental reason is that, with rare exceptions (see
\cite{Hasselblatt2007}), the sequence verifies a finite recurrence relation (with
integer coefficients).

This phenomenon has first been noticed for birational maps of finite
dimensional projective spaces, and extended to the infinite
dimensional case \cite{Tremblay2001,Viallet2006}.  This property can be proven in the
finite dimensional case in a number of instances, by a careful analysis of the
singularity structure of the map \cite{Takenawa2001JPhyA,GG_cremona3}. Another approach using
the stabilisation of the form of the iterates was developed for the
infinite dimensional case \cite{Viallet2015}. We will not give a complete
proof for the models we study here. We will rather use a more
heuristic {\em but extremely fast and efficient } method, which is to
{\em look for a generating function for the sequences of degrees}. If the
generating function happens to be rational, we will find it using
Pad\'e approximant, provided we are able to calculate sufficiently
many terms, as will be the case for all examples we consider here.

Given a sequence $\Set{d_k}_{k\in\N_0}$
its generating function is the function admitting the sequence as Taylor 
coefficients:
\begin{equation}
    g(z) = \sum_{k=0}^{\infty} d_{k} z^{k}.
    \label{eq:genfunc}
\end{equation}
  In a large number of instances (and this is the case for the examples that will be
  considered here) {\em the generating function for the sequence of
    degrees of iterates is rational}.

This has two immediate consequences:
\begin{itemize}[itemsep=2pt,topsep=5pt]
    \item The entropy is given by the inverse of the smallest modulus of the
        poles of $g(z)$, since the position of the smallest pole of $g$ governs
        the asymptotics of its  Taylor coefficients.  If $g(z)=P(z)/Q(z)$,
        \begin{equation}
           \mathcal{S}  = 
            -\log \min\Set{|z|\in\R^{+} \quad|\quad Q(z)=0}.
            \label{eq:algentgenfunc}
        \end{equation}
    \item The denominator gives a finite recurrence obeyed, at least after a
        finite number of steps,  by the sequence $\Set{ d_k}$. What happens in
        addition is that the coefficients of this recurrence relation \emph{are
        integers}.
\end{itemize}

If all the poles of $g= P/Q$ lie on the unit circle, the denominator $Q(z)$ factorises as:
\begin{equation}
    Q\left( z \right) = \left( 1-z \right)^{\beta_{0}} 
    \prod_{i=1}^{K} \left( 1-z^{\beta_{i}} \right),
    \quad
    \beta_i \in \N,
    \label{eq:factor}
\end{equation}
the growth of the degree sequence is polynomial, and {\em the entropy vanishes}. 

We recall some properties  of rational generating functions, obtained by
combining results from the books~\cite[Chap. 6]{Elaydi2005},~\cite[Sect.
4.4]{Goldberg1986book}, and~\cite{FlajoletOdlyzko1990}. For further discussion
on the subject and more references we refer to the
reviews~\cite{GubbiottiASIDE16, GrammaticosHalburdRamaniViallet2009}.

\begin{theorem}
    A sequence $\left\{ d_{k} \right\}_{k\in\N_{0}}$ admits a rational
    generating function $g\in\Cn\left( z \right)$ if and only if it 
    solves a \emph{linear difference equation with constant coefficients}.
    Moreover, if $\rho>0$ is the radius of convergence of $g$, writing
    $g$ as:
    \begin{equation}
        g = A\left( z \right) + B\left( z \right)\left( 1-\frac{z}{\rho} \right)^{-\beta},
        \quad \beta\in\N,
        \label{eq:gasy}
    \end{equation}
    where $A$ and $B$ are analytic functions for $\abs{z}<r$ such that
    $B(\rho)\neq 0$ we have:
    \begin{equation}
        d_{k} \sim \frac{B\left( \rho \right)}{\Gamma\left( \beta \right)}
        k^{\beta-1}\rho^{-k},
        \quad
        k \to \infty,
        \label{eq:dasy}
    \end{equation}
    where $\Gamma(z)$ is the Euler Gamma function.
    \label{th:rgf}
\end{theorem}

\begin{cor}
    With the  hypotheses of \Cref{th:rgf} and the additional assumption
    that $\rho=1$, then
    \begin{equation}
        d_{k} \sim \gamma \,  k^{\beta-1},
        \quad
        k \to \infty,
        \label{eq:dasyint}
    \end{equation}
    where $\beta=\beta_{0}+K$, with $\beta_0$ and $K$ as in  \Cref{eq:factor} and $\gamma$ a constant factor.
    \label{cor:int}
\end{cor}

In short, \Cref{th:rgf} and \Cref{cor:int} tell us that given a rational generating
function we can estimate the growth, which in the integrable case is
polynomial. In \Cref{sec:examples} we will make extensive use of these two results.

\subsubsection{Practical considerations}

It is not possible to compute  a large number of iterates for a
generic initial data, as this would imply manipulating huge formal
expressions in the $4N+1$ homogeneous coordinates of $\mathcal{X}=
\Pj^{4N}$ . The degrees we are looking for can equivalently be
obtained looking at the image of a randomly chosen generic straight line.

Consider  for example the case of the north-east evolution (see
\Cref{fig:indexing}). The initial data are $x_a, x_b, x_e, x_f$,  
for the cells of index $[1,j], j= 1\dots N$, and 
we take these  data on a straight line in $\mathcal{X}$:
\begin{equation}
    x_{v}[1,j]= \frac{\alpha_{{v},j} t + \beta_{{v},j}}{\alpha_{0} t + \beta_{0}},
    \quad
    {v}= a,b,e,f,\quad j=1,\dots,N,
    \label{eq:iniCP}
\end{equation}
where the $\alpha$'s and $\beta$'s  are fixed constants, and $t$ is the parameter on the line. The degree we analyse is the degree in $t$.

A final simplification is obtained by using polynomials over the
finite field $\mathbb{Z}_p$ (integers modulo some large prime $p$) to
perform the explicit calculations. An adequate choice of value for $p$
saves us from accidental drop of degrees due to undesirable extraneous
factorisations.


\subsection{Upper bound on the value of the algebraic entropy}

In this subsection we state and prove an upper bound on the value of
the algebraic entropy for a generic rotationally symmetric hex system. By {\it generic} we will mean a hex system defined by the
embedding \eqref{Ahexdef}, but we are not looking for specific
realisations in terms of face-centered quad equations satisfying CAFCC. That is,
for us the equation will be given by the following (formal) expression:
\begin{equation}
    \begin{aligned}
    A(x;x_\alpha,x_\beta,x_\gamma,x_\delta) &=\kappa_1x_\alpha x_\beta x_\gamma x_\delta  
    +\kappa_2x_\alpha x_\beta x_\gamma +\kappa_3x_\alpha x_\beta x_\delta  
    +\kappa_4x_\alpha x_\gamma x_\delta  +\kappa_5x_\beta x_\gamma x_\delta  
    \\
    &+\kappa_6x_\alpha x_\beta +\kappa_7x_\alpha x_\gamma +\kappa_8x_\alpha x_\delta  
    +\kappa_9x_\beta x_\gamma +\kappa_{10}x_\beta x_\delta  +\kappa_{11}x_\gamma x_\delta  
    \\
    &+\kappa_{12}x_\alpha+\kappa_{13}x_\beta +\kappa_{14}x_\gamma +\kappa_{15}x_\delta  +\kappa_{16}=0,
    \end{aligned}
    \label{eq:afflin}
\end{equation}
where $\kappa_i=\kappa_i(x)$ are polynomials of arbitrary degree.
The dependence on the constants $\alpha$, $\beta$, and $\gamma$ will not
be considered. Note that the expression \eqref{eq:afflin} is \emph{multilinear}
in $x_i$, $i\in\Set{\alpha,\beta,\gamma,\delta}$.


\begin{prop}
    Assume we are given a rotationally symmetric hex system
    $\boldsymbol{A}^{(hex)}$ as in equation~\eqref{eq:afflin}.  Recalling
    from equation~\eqref{typecsym} that $\mu$ is maximal degree in $x$ of the
    polynomials $\kappa_{i}=\kappa_{i}(x)$, we have that the global algebraic
    entropy is bounded by:
    \begin{equation}
        S_\text{Max}(\mu) = \log \left(1 + \sqrt{\mu+2}\right),
        \label{eq:aemax}
    \end{equation}
    regardless of the direction.
    \label{prop:generic}
\end{prop}

\begin{remark}
    \Cref{prop:generic} is the analogue for hex systems of the bound for finite
    dimensional maps~\cite[Sect 2.1]{GubbiottiASIDE16}, for quad
    equations~\cite[Sect. 3.6.1]{HietarintaBook}, and face-centered quad
    equations~\cite[Prop. 3.1]{GubKelsEntropy}.  Moreover, note that for
    $\mu=0$ in \Cref{eq:aemax}, i.e.\ the functions $\kappa_i$ are constants,
    we obtain again the generic growth for a multilinear quad equation:
    \begin{equation}
        S_\text{Max,quad} = \log \left(1 + \sqrt{2}\right).
        \label{eq:aemaxquad}
    \end{equation}
    Finally, we remark that the maximal entropy of hexagonal equations 
    as $\mu\to\infty$ is less of the one for face-centered quad equations,
    given by~\cite[Equation (2.23)]{GubKelsEntropy}:
    \begin{equation}
        S_\text{Max,fcq}(\mu) = \log \left[\frac{1}{2}\left(\mu+2 + \sqrt{\mu^2+4\mu+8}\right)\right],
        \label{eq:aemaxfcq}
    \end{equation}
    for all $\mu>0$. However, by taking the limit as $\mu\to\infty$ of the
    ratio of formulas~\eqref{eq:aemax} and~\eqref{eq:aemaxfcq} we see that they
    approach infinity at the same speed.
    \label{rem:genericgrowth}
\end{remark}

\begin{proof}
    First we note that since we want to establish an upper bound we can 
    pick up any direction of evolution we want. For sake of simplicity we 
    choose the direction NE. Then, we consider from \Cref{Ahexdef} and
    \eqref{eq:afflin} the two equations $A(x_f;x_e,x_b,x_c,x_a)=0$ and
    $A(x_a;x_b,x_e,x_d,x_f)=0$ and solve with respect to $x_c$ and $x_d$,
    see \Cref{tab:Dist}. We get two expressions of the following form:
    \begin{subequations}
        \begin{align}
            x_c &= \frac{P_c(x_f;x_a,x_b,x_e)}{Q_c(x_f;x_a,x_b,x_e)}
            \\
            x_d &= \frac{P_d(x_a;x_b,x_e,x_f)}{Q_d(x_a;x_b,x_e,x_f)},
        \end{align}
        \label{eq:xcdexpl}%
    \end{subequations}
    where $\deg_{x_a} P_c = \deg_{x_b} P_c = \deg_{x_e} P_c =1$, $\deg_{x_f}
    P_c = \mu$, $\deg_{x_a} P_d = \mu$, $\deg_{x_b} P_d = \deg_{x_e} P_d =
    \deg_{x_f} P_d =1$, and analogous formulas hold for the polynomials $Q_c$
    and $Q_d$. Note that the total degree of $P_c$, $P_d$, $Q_c$, and $Q_d$ is
    $\mu+3$. Then, assuming there is no factorisation and calling the degree
    with respect to a vertex $\delta_{v}$ for $v\in D_{v}$ we obtain the
    following equations:
    \begin{subequations}
        \begin{align}
            \delta_c &= \delta_a + \delta_b +  \delta_e + \mu \delta_f,
            \\
            \delta_d &= \mu \delta_a + \delta_b + \delta_e + \delta_f.
        \end{align}
        \label{eq:deltacdexpl}%
    \end{subequations}
    Through the non-factorisation assumption, in the case of considering the
    maximum, we can put $\delta_c=\delta_d := \delta^{(N+2)}$, 
    $\delta_e=\delta_f := \delta^{(N+1)}$, and $\delta_a=\delta_f := \delta^{(N)}$.
    The indexing follows from the fact that on a generic hexagon after
    the second iterates there are three different levels of iterations,
    see \Cref{fig:3levit}.
 
    Then, the two equations in~\eqref{eq:deltacdexpl} collapse into the
    single one:
    \begin{align}
        \delta^{(N+2)} &= 2\delta^{(N+1)} + (\mu+1) \delta^{(N)},
        \quad \forall N \geq 2.
        \label{eq:deltacdexplred}
    \end{align}
    \Cref{eq:deltacdexplred} is complemented with the initial conditions
    $\delta^{(0)}=1$, which follows by definition, and $\delta^{(1)}=\mu+3$.
    This second one follows from~\eqref{eq:xcdexpl}, since for instance:
    \begin{equation}
        \delta^{(1)} = \deg P_{c}(x_{f};x_{a},x_{b},x_{e}) 
        =
        \deg_{x_a} P_c + \deg_{x_b} P_c + \deg_{x_e} P_c + \deg_{x_f}
        =\mu+3.
        \label{eq:delta1}
    \end{equation}
    Note that, with the given assumptions, calculating $\delta^{(1)}$ with
    $Q_{c}$, $P_{d}$, or $Q_{d}$ gives the same result as \eqref{eq:delta1}.
    Then, we solve the \Cref{eq:deltacdexplred} with the generating function
    method \cite[Sect. 4.4]{Goldberg1986book}. After imposing the initial
    conditions we obtain that the function:
    \begin{equation}
        g_{\delta}(z) = \frac{1+(\mu+1)z}{1-2z -(\mu+1)z^2},
        \label{eq:gfmax}
    \end{equation}
    is the generating function of the sequence $\Set{\delta^{(n)}}$.
    Since the singularities of $g_{\delta}$ are:
    \begin{equation}
        z_\pm = -\frac{1\pm\sqrt{2+\mu}}{\mu+1},
    \end{equation}
    from \Cref{eq:algentgenfunc} we obtain \Cref{eq:aemax} because the root of smaller
    absolute value is $z_{-}$.
\end{proof}

\begin{remark}
    We remark that the computation of the generating function in
    \Cref{eq:gfmax} is \emph{exact} because we have the explicit expression of
    recursion relation of the sequence~\eqref{eq:deltacdexpl}. Note that in
    the non-integrable cases with maximal growth the form of the iterates
        stabilises \emph{after three iterations}. We conclude by
        observing that this result does not exclude the existence of
    non-integrable hex systems with slower rate of growth.
    \label{rem:gf}
\end{remark}



\section{Examples}
\label{sec:examples}

In this section we will give some explicit examples of the calculation of entropy  of hex
systems. In particular, we will examine the consistent systems given  in \cite{Kels2022},  which are all found to have vanishing entropy. We will also consider some systems 
of non-consistent equations and show that, as expected, they are non-integrable 
according to the algebraic entropy test. In particular we will confront their 
growth with the standard one presented in \Cref{prop:generic}.

\begin{remark}
    In \ref{app:equations} we list the explicit form of the face-centered
    quad equation we are going to use throughout this section.
    In the so-called hyperbolic cases, exponential and hyperbolic functions
    are present. Following the recipe given in \Cref{sec:aedef} it is
    preferable to use integer numbers. To do so we will make systematic
    use of the substitution:
    \begin{equation}
        \EXP^{\xi} = \tau \in \Z,
        \quad 
        \xi = \alpha,\beta,\gamma.
        \label{eq:intvalues}
    \end{equation}
    This is not restrictive, see \cite{GubbiottiASIDE16,GubKelsEntropy} 
    and also \cite{Viallet2009} where a similar substitution was used
    for elliptic parameters.
    \label{rem:params}
\end{remark}

\subsection{Hex systems of type A}

Because of their symmetries, the simplest type of hex systems to consider are the systems of type A, 
which are obtained from \eqref{Ahexdef}.  For the equations $A2_{(\delta_1;\delta_{2})}$, the evolution of the 
system can be written explicitly in the NE direction solving for 
$x_c$ and $x_d$, which are given by
\begin{subequations}
    \begin{align}
        x_c &=\frac{P_{(\delta_1;\delta_{2})}\bigl((x_a)^{1-\frac{\delta_2}{2}};x_f,x_e,x_b;\gm,\bt,\al\bigr)}{Q_{(\delta_1;\delta_{2})}\bigl((x_a)^{1-\frac{\delta_2}{2}};x_f,x_e,x_b;\gm,\bt,\al\bigr)},
        \\
        x_d &=\frac{P_{(\delta_1;\delta_{2})}\bigl((x_f)^{1-\frac{\delta_2}{2}};x_a,x_b,x_e;\al,\bt,\gm\bigr)}{Q_{(\delta_1;\delta_{2})}\bigl((x_f)^{1-\frac{\delta_2}{2}};x_a,x_b,x_e;\al,\bt,\gm\bigr)},
    \end{align}
    \label{eq:A2d1d2}
\end{subequations}
where
\begin{subequations}
    \begin{align}
        P_{(\delta_1;\delta_{2})}(x_f;x_a,x_b,x_e;\al,\bt,\gm) &=
        \begin{aligned}[t]
            &(\al-\gm + x_f)^{{\delta_2}+1}
            \bigl( (\bt + x_f)^{1+{\delta_2}} - x_a\bigr)\times
            \\
            &\phantom{\times}
            \bigl(( \gm -\bt + x_f)^{1+{\delta_2}} - x_b\bigr)
            \bigl((x_f-\al )^{1+{\delta_2}} - x_e\bigr)
            \\
            &+(-1)^{\delta_2} ( \al-\gm - x_f)^{{\delta_2}+1}
            \bigl(( x_f-\bt )^{1+{\delta_2}} - x_a\bigr)\times
            \\
            &\phantom{\times}\bigl( (\bt - \gm + x_f)^{1+{\delta_2}} - x_b\bigr)
            \bigl( (\al + x_f)^{1+{\delta_2}} - x_e\bigr)
            \\
            &-2(1-{\delta_1})\bigl((\bt-\al)(\al+\bt-\gm)\gm x_f 
            \\
            &+ (\al-\gm)(\gm-\bt)(\al x_a-\bt x_e) + \al \bt(\al-\gm)x_b\bigr),
        \end{aligned}
        \\
        Q_{(\delta_1;\delta_{2})}(x_f;x_a,x_b,x_e;\al,\bt,\gm) &=
        \begin{aligned}[t]
            &\bigl((\bt + x_f)^{1+{\delta_2}} - x_a\bigr) 
            \bigl(( \gm -\bt + x_f)^{1+{\delta_2}} - x_b\bigr)
            \times
            \\
            &\phantom{\times}\bigl((x_f-\al)^{1+{\delta_2}} - x_e\bigr) 
            \\
            &-\bigl((-\bt + x_f)^{1+{\delta_2}} - x_a\bigr) 
            \bigl((\bt - \gm + x_f)^{1+{\delta_2}} - x_b\bigr)\times
            \\
            &\phantom{\times}\bigl((\al + x_f)^{1+{\delta_2}} - x_e\bigr) 
            - 2(1-{\delta_1})\al \bt(\bt-\gm).
        \end{aligned}
    \end{align}
    \label{eq:PQd1d2}
\end{subequations}
Note that the expressions for $x_c$ and $x_d$ are exchanged by the substitutions
\begin{equation}\label{cdsubstitution}
x_a\leftrightarrow x_f,\quad x_b\leftrightarrow x_e,\quad \al\leftrightarrow\gm.
\end{equation}

The integrable cases are expected for $(\delta_1,\delta_2)=(0,0),(1,0),(1,1)$, which are known to be consistent.
Indeed, computing the growth of degrees for both vertices $x_{c}$ and $x_{d}$ 
for these values of the pair $(\delta_1,\delta_2)$, we obtain the same sequence.
Explicitly this is given by:
\begin{equation}
    1, 3, 7, 13, 21, 31, 43, 57, 73, 91, 111, 133, 157\ldots.
    \label{eq:A200growth}
\end{equation}
This degree sequence is fitted by the following generating function:
\begin{equation}
    g(z)= \frac{z^2+1}{(1-z)^3},
    \label{eq:A200gf}
\end{equation}
The denominator has $z=1$ as the only zero, giving \emph{zero entropy}. From
\Cref{cor:int} the growth is \emph{quadratic}, and using e.g.\ the 
$\mathcal{Z}$-transform \cite[Chap. 6]{Elaydi2005}, it can be shown that $d_n = n(n+1)+1$.
From the symmetries given in \Cref{sec:hex} we have that the same considerations apply 
to all the other directions of evolution.

For the equations $A3_{(\delta)}$, the hex system in the NE direction may be solved 
for
\begin{equation}
    x_c=
    \frac{\splitfrac{(x_e \shn{\al}{\bt} + x_f \shn{\bt}{} - x_b \shn{\gm}{})x_a^2 - x_a x_e x_f \shn{\al}{} + x_b x_e x_f \shn{\al}{\gm} - x_a x_b x_e\shn{\phi}{} - x_a x_b x_f \shn{\bt}{\gm} }{+
    \delta ( x_a \shn{\phi}{}\shn{\bt}{\gm} \shn{\al}{} + \shn{\al}{\gm} (x_e \shn{\bt}{} \shn{\gm}{} + \shn{\al}{\bt} (x_f \shn{\gm}{} - x_b \shn{\bt}{}))}}
{\delta \shn{\al}{\bt}\shn{\bt}{}\shn{\gm}{} - x_a x_b \shn{\al}{} + x_b x_f\shn{\al}{\bt} + x_b x_e \shn{\bt}{} + x_a^2 \shn{\al}{\gm} - x_a x_f\shn{\phi} - x_a x_e\shn{\bt}{\gm} - x_e x_f\shn{\gm}{}},
    \label{eq:A31c}
\end{equation}
where
\begin{equation}
\shn{\al}{}=\sinh(\al),\qquad\shn{\al}{\bt}=\sinh(\al-\bt),
\end{equation}
and
\begin{equation}
\phi=\al+\bt-\gm,
\end{equation}
and the expression for $x_d$ is obtained from $x_c$ by the substitution \eqref{cdsubstitution}.
For $\delta=0,1$, the growth for both vertices $x_{c}$ and $x_{d}$ is 
given by the sequence \eqref{eq:A200growth}, and therefore fitted by the 
generating function \eqref{eq:A200gf}. That is, also for the hex system 
generated by $A3_{(\delta)}$, $\delta=0,1$, we have the quadratic growth 
$d_{n}=n(n+1)+1$. Again, by the symmetries given in \Cref{sec:hex} we have that the same 
considerations apply to all the other directions of evolution.

\subsection{Symmetric systems of type C}

The hex systems of type C where $C=\overline{C}$ in~\eqref{Chexdef} will be
referred to as symmetric systems. Despite the underlying face-centered equation
being of type C, their entropy properties as hex systems will be similar to the
properties observed above for hex systems of type A.  In this case, there are
four examples that are known to be consistent, given by: $C1_{(0)}$
(algebraic), $C2_{(1;0;0)}$ (rational), $C3_{(0;0;0)}$ (hyperbolic), and
$C3_{(1;0;0)}$ (hyperbolic).

As expected from their face-centered analogues these equations display a
much wider variety of behaviour than the systems of type A. However, as expected from
their consistency property they all possess a quadratic growth in all
directions and in all vertices, making them integrable in the sense of 
the algebraic entropy. In total, we identified eight new degree sequences,
which are displayed in \Cref{tab:seqsym}, where the disposition of the sequences
in the various directions and vertices is shown in \Cref{tab:evolsym}.

\begin{table}[tbh]
    \centering
    \begin{tabular}{cl}
        \toprule
        \textbf{Label} & \multicolumn{1}{c}{\textbf{Degree sequence}}
        \\
        \midrule
        $s_{1}$ & 1, 3, 7, 9, 13, 19, 23, 29, 37, 43, 51, 61, 69, 79, 91, 101, 113\dots
        \\
        $s_{2}$ & 1, 3, 5, 7, 11, 15, 19, 25, 31, 37, 45, 53, 61, 71, 81, 91, 103\dots
        \\
        $s_{3}$ & 1, 3, 7, 11, 17, 23, 29, 37, 45, 53, 63, 73, 83, 95, 107, 119, 133\dots
        \\
        $s_{4}$ & 1, 3, 6, 10, 15, 20, 26, 33, 40, 48, 57, 66, 76, 87, 98, 110, 123\dots
        \\
        $s_{5}$ & 1, 3, 7, 10, 14, 20, 25, 31, 39, 46, 54, 64, 73, 83, 95, 106, 118\dots
        \\
        $s_{6}$ & 1, 3, 6, 8, 12, 17, 21, 27, 34, 40, 48, 57, 65, 75, 86, 96, 108\dots
        \\
        $s_{7}$ & 1, 3, 7, 12, 18, 24, 31, 39, 47, 56, 66, 76, 87, 99, 111, 124, 138\dots
        \\
        $s_{8}$ & 1, 3, 7, 11, 16, 22, 28, 35, 43, 51, 60, 70, 80, 91, 103, 115, 128\dots
        \\
        \bottomrule
    \end{tabular}
    \caption{Sequences of degree for the symmetric hex systems of type C}
    \label{tab:seqsym}
\end{table}

\begin{table}[tbh]
    \centering
    \begin{tabular}{cccccccc}
        \toprule
        \multirow{3}*{\textbf{System}} & \multicolumn{6}{c}{\textbf{Direction}}
        \\
        & \multicolumn{2}{c}{NE} & \multicolumn{2}{c}{N} &\multicolumn{2}{c}{NW}
        \\
        & $x_{c}$ & $x_{d}$ & $x_{b}$ & $x_{c}$ &$x_{a}$ & $x_{b}$
        \\
        \midrule
        $C1_{0}$
        & $s_{1}$ & $s_{2}$ & $s_{2}$ & $s_{1}$ & 
        \eqref{eq:A200growth} & \eqref{eq:A200growth}
        \\
        $C2_{(1;0;0)}$
        & $s_{3}$ & $s_{4}$ & $s_{4}$ & $s_{3}$ & 
        \eqref{eq:A200growth} & \eqref{eq:A200growth}
        \\
        $C3_{(0;0;0)}$
        & $s_{5}$ & $s_{6}$ & $s_{6}$ & $s_{5}$ & 
        \eqref{eq:A200growth} & \eqref{eq:A200growth}
        \\
        $C3_{(1;0;0)}$
        & $s_{7}$ & $s_{8}$ & $s_{8}$ & $s_{7}$ & 
        \eqref{eq:A200growth} & \eqref{eq:A200growth}
        \\
        \bottomrule
    \end{tabular}
    \caption{Sequences in the various direction of evolution
    for the symmetric hex systems of type C.}
    \label{tab:evolsym}
\end{table}

 In particular note that:
\begin{itemize}[itemsep=2pt,topsep=5pt]
    \item the sequences of degrees in the NE and N directions are different
        for the different equations;
    \item in the NW direction the evolution is the same for the
        two vertices, and it is same for each of the four systems:
        it is given by the sequence \eqref{eq:A200growth} with
        generating function \eqref{eq:A200gf}, and explicit form
        $d_n = n(n+1)+1$;
    \item for the evolution in the NE and N direction the generating
        the generating function can be written as:
        \begin{equation}
            g(z) = \frac{\nu(z)}{(z-1)^{3}(z^{2}+z+1)},
            \label{eq:gfasymm}
        \end{equation}
        with numerator dependent from the sequence, given in \Cref{tab:numsymm};
    \item given the structure of the evolution we have that the
        maximal growth coincides, either with the sequence $s_{i}$ with $i=1,3,5,7$,
        or \eqref{eq:A200growth}.
\end{itemize}

\begin{table}[tbh]
    \centering
    \begin{tabular}{cc}
        \toprule
        \textbf{Label} & \textbf{Numerator} $\nu(z)$
        \\
        \midrule
        $s_{1}$ & $-(z^4-3 z^3+2 z^2+z+1)$
        \\
        $s_{2}$ & $-(z^4-z^3+z+1)$
        \\
        $s_{3}$ & $(z^2+1) (2 z^3-z^2-z-1)$
        \\
        $s_{4}$ & $(z^2 + 1) (z^3 - z - 1)$
        \\
        $s_{5}$ & $(2 z^3 - 2 z^2 - z - 1)$
        \\
        $s_{6}$ & $-(z^4 - 2 z^3 + z^2 + z + 1)$
        \\
        $s_{7}$ & $2 z^5-2 z^2-z-1$
        \\
        $s_{8}$ & $z^5 + z^3 - 2 z^2 - z - 1$
        \\
        \bottomrule
    \end{tabular}
    \caption{Numerators of the generating functions \eqref{eq:gfasymm}.}
    \label{tab:numsymm}
\end{table}

\subsection{Asymmetric hex systems of type C}

The hex systems of type C where $C\neq \overline{C}$ will be referred to as asymmetric systems.  In this case, 
there are three examples that are known to be consistent, which are given by the pairs of equations: $\Set{C1_{(1)},C2_{(0;0;0)}}$,$\Set{C2_{(1;1;0)},C2_{(1;0;1)}}$, and 
$\Set{C3_{(1/2;1/2;0)}, C2_{(1/2;0;1/2)}}$.  
The asymmetric hex systems of type C that are constructed from these equations are all found to 
possess a quadratic growth in all
directions and in all vertices, making them integrable in the sense of 
the algebraic entropy. 

In total, we identified four new degree sequences,
which are displayed in \Cref{tab:seqasym}, with the disposition of the sequences
in the various directions and vertices is shown in \Cref{tab:evolasym}.

\begin{table}[tbh]
    \centering
    \begin{tabular}{cl}
        \toprule
        \textbf{Label} & \multicolumn{1}{c}{\textbf{Degree sequence}}
        \\
        \midrule
        $a_{1}$ & 1, 3, 7, 10, 14, 20, 25, 31, 39, 46, 54, 64, 73, 83, 95, 106\dots
        \\
        $a_{2}$ & 1, 3, 6, 8, 12, 17, 21, 27, 34, 40, 48, 57, 65, 75, 86, 96\ldots
        \\
        $a_{3}$ &1, 3, 7, 13, 21, 30, 40, 50, 61, 73, 85, 98, 112, 126, 141, 157\dots
        \\
        $a_{4}$ &1, 3, 7, 13, 21, 29, 38, 48, 58, 69, 81, 93, 106, 120, 134, 149\dots
        \\
        \bottomrule
    \end{tabular}
    \caption{Sequences of degree for asymmetric hex systems of type C}
    \label{tab:seqasym}
\end{table}

\begin{table}[tbh]
    \centering
    \begin{tabular}{cccccccc}
        \toprule
        \multirow{3}*{\textbf{System}} & \multicolumn{6}{c}{\textbf{Direction}}
        \\
        & \multicolumn{2}{c}{NE} & \multicolumn{2}{c}{N} &\multicolumn{2}{c}{NW}
        \\
        & $x_{c}$ & $x_{d}$ & $x_{b}$ & $x_{c}$ &$x_{a}$ & $x_{b}$
        \\
        \midrule
        $\Set{C1_{1},C2_{(0;0;0)}}$
        & $a_{1}$ & $a_{2}$ & $a_{2}$ & $a_{1}$ & 
        \eqref{eq:A200growth} & \eqref{eq:A200growth}
        \\
        $\Set{C2_{(1;1;0)},C2_{(1;0;1)}}$
        & $a_{3}$ & $a_{4}$ & $a_{4}$ & $a_{3}$ & 
        \eqref{eq:A200growth} & \eqref{eq:A200growth}
        \\
        $\Set{C3_{(1/2;1/2;0)}, C2_{(1/2;0;1/2)}}$
        & $a_{3}$ & $a_{4}$ & $a_{4}$ & $a_{3}$ & 
        \eqref{eq:A200growth} & \eqref{eq:A200growth}
        \\
        \bottomrule
    \end{tabular}
    \caption{Sequences in the various direction of evolution
    for the asymmetric hex systems of type C.}
    \label{tab:evolasym}
\end{table}

In particular note that:
\begin{itemize}[itemsep=2pt,topsep=5pt]
    \item the systems $\Set{C2_{(1;1;0)},C2_{(1;0;1)}}$ and 
        $\Set{C3_{(1/2;1/2;0)}, C2_{(1/2;0;1/2)}}$ share the same
        growth properties;
    \item in the NW direction the evolution is the same for the
        two vertices, and it is same for all the three systems:
        it is given by the sequence \eqref{eq:A200growth} with
        generating function \eqref{eq:A200gf}, and explicit form
        $d_n = n(n+1)+1$;
    \item for the evolution in the NE and N direction the generating
        the generating function can be written as in \Cref{eq:gfasymm}
        with numerator depending on the sequence, see \Cref{tab:numasymm};
    \item following \Cref{cor:int} in all cases the growth is quadratic;
    \item given the structure of the evolution we have that the
        maximal growth coincides, either with the sequence $a_{1}$, $a_{3}$,
        or \eqref{eq:A200growth}.
\end{itemize}

\begin{table}[tbh]
    \centering
    \begin{tabular}{cc}
        \toprule
        \textbf{Label} & \textbf{Numerator} $\nu(z)$
        \\
        \midrule
        $a_{1}$ & $2 z^3-2 z^2-z-1$
        \\
        $a_{2}$ & $-(z^4-2 z^3+z^2+z+1)$
        \\
        $a_{3}$ & $2 z^7+z^6+z^5-z^4-z^3-2 z^2-z-1$
        \\
        $a_{4}$ & $z^7+z^6+2 z^5-z^4-z^3-2 z^2-z-1$
        \\
        \bottomrule
    \end{tabular}
    \caption{Numerators of the generating functions~\eqref{eq:gfasymm}.}
    \label{tab:numasymm}
\end{table}

\subsection{Type B examples}

Type B face-centered quad equations satisfy the first two symmetries of
\eqref{fcqsyms}, and thus they are more symmetric than type C equations but
less symmetric than type A equations. The type B equations were previously
shown to be integrable in the square lattice as they have vanishing
algebraic entropy with linear growth of degrees \cite{GubKelsEntropy}.
However, it is not known how to construct consistent systems of type B
equations in the sense described in \Cref{sec:hex}.

One might attempt to construct a hex system of type B equations by simply
replacing type C equations in \eqref{Chexdef} type B
equations, but this does not result in a consistent hex system.  This means
that the evolution depends on which equations are chosen to solve for two
vertices in terms of the other four.  If $\B{x}{x_a}{x_b}{x_c}{x_d}{\al}{\bt}{\gm}=0$ denotes a type B
face-centered quad equation, such a hex system is given by
\begin{equation}\label{Bhexdef}
\begin{alignedat}{2}
(\Ba)&&\qquad \B{\xa}{\xe}{\xc}{\xb}{\xf}{\gm}{\bt}{\al}=0, \\
(\Bb)&&\qquad \B{\xb}{\xf}{\xd}{\xa}{\xe}{\gm}{\al}{\bt}=0, \\
(\Bf)&&\qquad \B{\xf}{\xb}{\xd}{\xa}{\xc}{\gm-\al}{-\al}{\bt-\al}=0, \\
(\Bd)&&\qquad \B{\xd}{\xf}{\xb}{\xc}{\xe}{\gm}{\bt}{\al}=0, \\
(\Bc)&&\qquad \B{\xc}{\xe}{\xa}{\xd}{\xf}{\gm}{\al}{\bt}=0, \\
(\Be)&&\qquad \B{\xe}{\xc}{\xa}{\xd}{\xb}{\gm-\al}{-\al}{\bt-\al}=0.
\end{alignedat}
\end{equation}
as an analogue of the type C hex system in \eqref{Chexdef}.

Because the above system is not consistent on the hexagon, for the
evolution in the NE direction there are seven different ways to solve for
the two variables $x_c$ and $x_d$ in terms of the four variables
$x_a,x_b,x_e,x_f$. First, either the equation $(\Ba)$ can be used to solve
for $x_c$, or the equation $(\Be)$ can be used to solve for $x_d$.  Then,
depending on which variable was determined, one either has the four
equations $(\Bb),(\Bf),(\Bc),(\Be)$ to solve for the variable $x_d$, or the
four equations $(\Ba),(\Bb),(\Bd),(\Be)$ to solve for the variable $x_c$.

Testing such a type B system of hex equations for the example of $B2_{(1;0;0)}$, using the
equation $(\Ba)$ to determine $x_c$ and the equation $(\Be)$ to determine
$x_d$, gives the following solutions for $x_c$ and $x_d$ for evolution in
the NE direction
\begin{subequations}
    \begin{align}
x_c=\frac{ \bt x_b + \gm(x_e-x_f) + \al(\gm-\bt+x_f-x_b) - x_b(x_a+x_e) - x_a(x_e-x_f)}{\bt - x_a - x_f}, \\
x_d=\frac{ \bt x_e + \al(x_b-x_a) + \gm(\bt-\al+x_a-x_e) + x_f(x_e-x_a) + x_b (x_e + x_f)}{\bt + x_a + x_f}.    
    \end{align}
    \label{eq:xcxdB2100}
\end{subequations}
The above evolution gives the following sequence of degrees
\begin{subequations}
    \begin{align}
        x_c:& 1, 2,5,12,30,74,184,456,1132,2808,6968\dots
        \\
        x_d:& 1, 3,7,18,44,110,272,676,1676,4160,10320\dots
    \end{align}
    \label{eq:xcxdB2100deg}
\end{subequations}
with generating functions
\begin{equation}
    g_v(z)= -\frac{\nu_v(z)}{2z^3-2z^2-2z+1},
    \quad
    v=c,d,
\end{equation}
with 
\begin{equation}
    \nu_v(z)=
    \begin{cases}
        (z-1)(z+1) & \text{if $v=c$},
        \\
        z^2-z-1 & \text{if $v=d$}.    
    \end{cases}
\end{equation}
The type B system of hex equations constructed this way has a non-vanishing
algebraic entropy.  Computing the inverse of the map \eqref{eq:xcxdB2100},
i.e. solving with respect to the vertices $x_a$ and $x_f$, we find the same
degree sequences as in \eqref{eq:xcxdB2100deg}.

The same calculation can be repeated for the other possible combinations of
(non-consistent) equations. The results are shown in
\Cref{tab:typeB,tab:typeBseq} and display that all these combinations
possess exponential growth. Therefore, we conclude that one cannot have an integrable evolution using the above hex system of type B, with the equation $B2_{(1;0;0)}$.  Similar results are expected for other examples of type B equations, which are not consistent.

\begin{table}[htb!]
    \centering
    \begin{tabular}{ccc}
        \toprule
        \textbf{Combination} & \multicolumn{2}{c}{\textbf{Degree sequence}}
        \\
        & $x_{c}$ & $x_{d}$
        \\
        \midrule
        $(\Ba),(\Bf)$ & $b_{1}$ & $b_{1}$
        \\
        $(\Ba),(\Bb)$
        & $b_{2}$ & $b_{3}$
        \\
        $(\Ba),(\Bc)$ & $b_{4}$ & $b_{5}$
        \\
        $(\Ba),(\Be)$ & $b_{6}$ & $b_{7}$
        \\
        $(\Bf),(\Bb)$ & $b_{7}$ & $b_{6}$
        \\
        $(\Bf),(\Bd)$ & $b_{5}$ & $b_{4}$
        \\
        $(\Bf),(\Be)$ & $b_{3}$ & $b_{2}$
        \\
        \bottomrule
    \end{tabular}
    \caption{Sequences of degree for the equation $B2_{(1;0;0)}$, using the
    various combinations of equations.}
    \label{tab:typeB}
\end{table}

\begin{table}[htb]
    \centering
    \begin{tabular}{ccc}
        \toprule
        \textbf{Label} & \textbf{Degree sequence} & \textbf{Generating
        function}
        \\
        $b_{1}$ & 1, 2,4,9,19,42,90,197,425,926,2004\dots
        &
        $\displaystyle\frac{-z^2+z+1}{z^3-3z^2-z+1}$
        \\[10pt]
        $b_{2}$ &1, 2,5,13,35,95,259,707,1931,5275,14411\dots
        &$\displaystyle\frac{-(z^2+z-1)}{(z-1)(2z^2+2z-1)}$
        \\[10pt]
        $b_{3}$ & 1, 3,8,22,60,164,448,1224,3344,9136,24960\dots
        &$\displaystyle\frac{1+z}{-2z^2-2z+1}$
        \\[10pt]
        $b_{4}$ & 1, 2,5,15,43,128,375,1108,3261,9616,28330\dots
        &$\displaystyle\frac{-(z^2+z-1)(z^2-z-1)}{z^5-2z^4-3z^3+4z^2+2z-1}$
        \\[10pt]
        $b_{5}$ & 1, 3,9,27,80,236,696,2051,6045,17814,52499\dots
        &$\displaystyle\frac{-(1+z)(z^3-z^2+1)}{z^5-2z^4-3z^3+4z^2+2z-1}$
        \\[10pt]
        $b_{6}$ &1, 2,5,12,30,74,184,456,1132,2808,6968\dots
        &$\displaystyle\frac{-(z-1)(1+z)}{2z^3-2z^2-2z+1}$
        \\[10pt]
        $b_{7}$ & 1, 3,7,18,44,110,272,676,1676,4160,10320\dots
        &$\displaystyle\frac{-(z^2-z-1)}{2z^3-2z^2-2z+1}$
        \\
        \bottomrule
    \end{tabular}
    \caption{Explicit form of the sequences and their generating functions.}
    \label{tab:typeBseq}
\end{table}

\section{Conclusions and outlook}
\label{sec:concl}


In this paper, the concept of algebraic entropy has been extended to hex
systems. These are systems of equations defined on hexagons, where consistency allows to define birational evolutions in each direction of the
honeycomb lattice starting from staircases of initial conditions, as shown in
detail in \Cref{sec:aedef}. The basic idea was to characterise the
integrability of such hex systems by the vanishing of the
entropy~\cite{Veselov1992,FalquiViallet1993}, which is a global index of
complexity~\cite{Arnold1990moser,Arnold1990}, as was previously done for quad
equations~\cite{Tremblay2001,Viallet2006,GSL_general,Gubbiotti2023DecioMemorial}.

Our notion of algebraic entropy was used to examine specific examples of
consistent hex systems constructed from face-centered quad equations
\cite{Kels2022}. We have exhibited --in relation to the notion of
consistency-- integrable cases and non-integrable cases. In particular,
integrable cases have quadratic growth. Up to now we did not find any
example with linear growth, differently from quad \cite{GSL_general} and
face-centered quad equations \cite{GubKelsEntropy}.

A number of interesting open questions remain to be investigated. Examples include:
\begin{itemize}[itemsep=2pt,topsep=5pt]
    \item Classification of hex systems.
    \item Generalisation to multi-component hex systems e.g.\ as was done for a quad
        equation in~\cite{Kels2019Z}.
    \item Find interpolating families, in the spirit of what
        was done for quad equations, e.g.\ the $Q_{V}$~\cite{Viallet2009} and
        the two-periodic $Q_{V}$ equation~\cite{GSL_QV}.
    \item Investigate analogues of hex systems defined on lattices with different cells
        (e.g.\ squares, octagons). 
   \item Discrete soliton solutions for hex systems.
    \item Continuous limits of hex systems that could possibly lead to
        partial differential or differential-difference systems of equations.
        The integrable cases will be of special interest.
\end{itemize}

\section*{Acknowledgements}

The research of GG was partially supported by the GNFM through Progetto Giovani
GNFM 2023: ``Strutture variazionali e applicazioni delle equazioni alle
differenze ordinarie'', CUP\_E53C22001930001. GG also acknowledges the kind
hospitality of LPTHE where part of this manuscript was written. 
The research of APK was partially supported through ARC Discovery Project
DP200102118.

\appendix

\renewcommand{\thesection}{Appendix \Alph{section}}
\renewcommand{\thesubsection}{\Alph{section}.\arabic{subsection}}
\renewcommand{\theequation}{\Alph{section}.\arabic{equation}}

\section{Examples of face-centered quad equations}\label{app:equations}

In the following, $\phi$ is defined by $\phi=\al+\bt-\gm$, and $\zz{\alpha}$ and $\sh{\alpha}$ are defined by
\begin{equation}\nonumber
\zz{\alpha}=\EXP^{\alpha},\qquad \sh{\alpha}=\sinh(\alpha).
\end{equation}


Face-centered quad equations are 5-point difference equations which satisfy a property called consistency-around-a-face-centered-cube \cite{Kels2019}. The following are multilinear polynomials for different type A face-centered quad equations (see Section \ref{sec:hex}).  They are each degree 2 in $x$.

\begin{equation}\nonumber
\begin{alignedat}{2}
A3_{(\delta)}\quad&&
\bigl(\sh{\al-\bt}(x_ax_c-x_bx_d) - \sh{\gm}(x_ax_b-x_cx_d) - \sh{\phi}(x_ax_d-x_bx_c)\bigr)x \\&&\quad
+\sh{\bt}(x_ax^2-x_bx_cx_d) - \sh{\bt-\gm}(x_bx^2-x_ax_cx_d) - \sh{\al}(x_cx^2-x_ax_bx_d) \\&&+ \sh{\al-\gm}(x_dx^2-x_ax_bx_c)  + \delta\Bigl(
\sh{\bt-\gm}\sh{\al-\gm}(\sh{\al}x_a-\sh{\bt}x_c) \\&&+
 \sh{\al}\sh{\bt}(\sh{\gm-\al}x_b-\sh{\gm-\bt}x_d) +
 \sh{\gm}\sh{\al-\bt}\sh{\phi}x
\Bigr)
\end{alignedat}
\end{equation}

\begin{equation}\nonumber
\begin{alignedat}{2}
 A2_{(\delta_1;\,\delta_2)}\quad&&
\gm(x_b-x_d)\bigl((x-x_a)(x-x_c)-\delta_1\al\bt(2x+x_a+x_c-(\al-\bt)^2)^{\delta_2}\bigr) \\ &&
-\al(x_c-x_d)\bigl((x-x_a)(x-x_b)-\delta_1\bt^2(2x+x_a+x_b-\bt^2)^{\delta_2}\bigr) \\ &&+
\bt(x_a-x_b)\bigl((x-x_c)(x-x_d)-\delta_1\al^2(2x+x_c+x_d-\al^2)^{\delta_2}\bigr) \\ &&
\delta_1\gm\phi\Bigl( 
 \bigl(\al x_a-\bt x_c+\delta_2(\al-\bt)\al\bt\bigr)\bigl(2x+x_b+x_d + (\al-\gm)(\gm-\bt)\bigr)^{\delta_2} \\ &&-
 x(\al-\bt)\bigl(x+x_a+x_b+x_c+x_d + \gm\phi-(\al-\bt)^2 - x_ax_cx^{-1}\bigr)^{\delta_2}\Bigr)  \\ &&
-\delta_2\gm\bigl(\phi(x_a\al^3-x_c\bt^3)-\al^2\bt^2(x_b-x_d)\bigr)
\end{alignedat}
\end{equation}

The following are multilinear polynomials for different type C face-centered quad equations.  They are each degree 2 in $x$.

\begin{equation}\nonumber
\begin{alignedat}{2}
C3_{(\delta_1;\,\delta_2;\,\delta_3)} && 
\Bigl(\zz{\gamma} x_d-x_c - \delta_3\zz{\alpha}\bigl(\zz{-\beta}(\zz{\gamma} x_b-x_a) + \zz{\bt-\gm}(\zz{\gamma} x_a - x_b)\bigr)\Bigr)x^2 \\ && +
(x_d-\zz{\gamma} x_c)\bigl(x_ax_b+2\delta_2\sh{\bt}\sh{\bt-\gm}\bigr) +\Bigl(\zz{\gm-\bt}(x_bx_c - x_ax_d) \phantom{x}\\ && + 
\zz{\beta}(x_ax_c - x_bx_d) +
  \bigl(\delta_1\zz{-\alpha}+\zz{\al-\gm}(\delta_2x_cx_d-\delta_3x_ax_b)\bigr) 
  \bigl(1-\zz{2\gamma}\bigr)\Bigr)x  \\ && +
 \delta_1\zz{-\alpha}\bigl(\zz{\beta}(\zz{\gamma} x_b-x_a) 
 + \zz{\gm-\bt}(\zz{\gamma} x_a - x_b)\bigr) 
 \bigl(1+2\delta_2\zz{2\al-\gm}x_cx_d\bigr)\phantom{x}
\end{alignedat}
\end{equation}

\begin{equation}\nonumber
\begin{alignedat}{2}
 C2_{(\delta_1;\,\delta_2;\,\delta_3)} &&\quad 
\bigl(\gm(x-x_a)(x+x_b)+2\bt x(x_a-x_b)\bigr)(\gm-2\al)^{\delta_3}  
+  (x_c-x_d)\bigl(x(x_a+x_b-x)-x_ax_b\bigr) \\ &&+
2\delta_2\bigl(\al(\al-\gm)-x_cx_d\bigr)\bigl(\gm(x-x_a)+\bt(x_a-x_b)\bigr) \\ && + 
2\delta_2\bt\gm(\bt-\gm)\bigl(x(x_c-x_d)\gm^{-1}+ \al(\al-\gm)+x_cx_d+\al(x_c+x_d)-\gm x_c\bigr) \\ && +
\delta_1   \bigl(\gm(x-x_a)+\bt(x_a-x_b)\bigr)\bigl((2\al-\gm)(\gm-2\al)^{\delta_3}+x_c+x_d\bigr)(\gm-2\al)^{\delta_2}   \\ && +
   \bt(\bt-\gm)\bigl(x_c-x_d+\gm(\gm-2\al)^{\delta_3}\bigr)(\bt(\gm-\bt)+x_a+x_b)^{\delta_2}  \\ && +
2\delta_3\Bigl( \bt(x_b-x_a)\bigl(x^2-\al(\al-\gm)\bigr) + \gm xx_b(x_a-x) + \gm(\al-\bt)\phi x - \al(\al-\gm)\gm x_a\Bigr)
\end{alignedat}
\end{equation}

\begin{equation}\nonumber
\begin{alignedat}{2}
C1_{(\delta)} && \quad
(x_c-x_d)\bigl(x^2-(x_a+x_b)x+x_ax_b-\delta\bt(\bt-\gm)\bigr)  +
 2\bigl(\bt(x_a-x_b)+\gm(x-x_a)\bigr)\bigl(-\tfrac{x_c+x_d}{2}\bigr)^{\delta}
\end{alignedat}
\end{equation}

The following multilinear polynomials for type B face-centered quad equations.  They are each degree 2 in $x$.

\begin{equation}\nonumber
\begin{alignedat}{2}
B3_{(\delta_1;\,\delta_2;\,\delta_3)} &&\quad \bigl(x_bx_c-x_ax_d + \tfrac{\delta_3}{2}(\tfrac{\al}{\bt}-\tfrac{\bt}{\al})(\gm-\tfrac{1}{\gm})\bigr)x \\ &&+
\delta_3\bigl((\gm x_a - x_b)\tfrac{1}{\al} - (\gm x_c - x_d)\tfrac{1}{\bt}\bigr)x^2 \\ &&+
\delta_1\bigl((\tfrac{x_a}{\gm}-x_b)(\al+2\delta_2x_cx_d\tfrac{\gm}{\al})-(\tfrac{x_c}{\gm}-x_d)(\bt+2\delta_2x_ax_b\tfrac{\gm}{\bt})\bigr)
\end{alignedat}
\end{equation}

\begin{equation}\nonumber
\begin{alignedat}{2}
B2_{(\delta_1;\,\delta_2;\,\delta_3)} && (x_ax_d-x_bx_c)\phi^{\delta_3}+\delta_1\bigl(x(-x)^{\delta_2}+(\delta_2+\delta_3)\al\bt\bigr)(x_a-x_b-x_c+x_d)  \\ &&+
\delta_1\bigl(\al(x_b-x_a) + \gm(x_a-x_c) + \bt(x_c-x_d) + \gm(\al-\bt)(-2x)^{\delta_2}\bigr)(-2x+\phi)^{\delta_2}\phi^{\delta_3} \\ &&+
\delta_2\Bigl(\gm\bigl(\al(x_b+x_c)-\bt(x_a+x_d)\bigr) + \gm(\al-\bt)\bigl(2\al\bt-\gm(\al+\bt)-2x^2\bigr) \Bigr) \\ &&+
\delta_3\Bigl((\al-\bt)(x_ax_c-x_bx_d) - \gm(x_ax_b-x_cx_d) \\ &&+
(x_a+x_d)(x_bx_c-\al\gm) - (x_b+x_c)(x_ax_d-\bt\gm)\Bigr)
\end{alignedat}
\end{equation}

\printbibliography

\end{document}